\documentclass[aps,pra,floatfix,twocolumn,nofootinbib,showpacs]{revtex4-1}

\usepackage{graphicx}
\usepackage{amssymb,amsmath,amsfonts,mathrsfs,eufrak,fancyhdr}
\usepackage{braket}

\usepackage{epsf}
\usepackage{color}

\def\sH{\mathscr{H}}
\def\Tr{\operatorname{Tr}}
\def\e{\mathrm{e}}

\definecolor{darkgreen}{RGB}{0, 150, 0}
\newcommand{\GREEN}[1]{{\color{darkgreen}#1}}
\newtheorem{corollary}{Corollary}[section]
\newtheorem{theorem}{Theorem}[section]

\begin{document}
\title{Percolated quantum walks with a general shift operator: From trapping to transport}
\author{J. Mare\v s, J. Novotn\'y, I. Jex}
\affiliation{Department of Physics, Faculty of Nuclear Sciences and Physical Engineering, Czech Technical University in Prague, B\v rehov\'a 7, 115 19 Praha 1 - Star\'e M\v esto, Czech Republic}
\date{\today}
\begin{abstract}
We present a generalized definition of discrete-time quantum walks convenient for capturing a rather broad spectrum of walker's behavior on arbitrary graphs. It includes and covers both: the geometry of possible walker's positions with interconnecting links and the prescribed rule in which directions the walker will move at each vertex. While the former allows for the analysis of inhomogeneous quantum walks on graphs with vertices of varying degree, the latter offers us to choose, investigate, and compare quantum walks with different shift operators. The synthesis of both key ingredients constitutes a well-suited playground for analyzing percolated quantum walks on a quite general class of graphs. Analytical treatment of the asymptotic behavior of percolated quantum walks is presented and worked out in details for the Grover walk on graphs with maximal degree $3$. We find, that for these walks with cyclic shift operators the existence of an edge-3-coloring of the graph allows for non-stationary asymptotic behavior of the walk. For different shift operators the general structure of localized attractors is investigated, which determines the overall efficiency of a source-to-sink quantum transport across a dynamically changing medium. As a simple nontrivial example of the theory we treat a single excitation transport on a percolated cube.
\end{abstract}

\maketitle

\section{Introduction}

Quantum walks are a rather popular model in several branches of
modern physics. Since its introduction \cite{quantum_random_walks,Meyer} they are the subject of numerous studies as well as
experiments \cite{experiments} of high sophistication. Quantum walks
are interesting by themselves \cite{review} but this interest is
further boosted by the possibly wide spectrum of applications ranging
from quantum information processing to simulating coherent quantum
transport in designed media. In this respect quantum walks follow
the same path as (classical) random walks which became a "classic"
in the field of statistical mechanics and are applied to a wide
range of problems starting from diffusion and ending with
description of economic trends \cite{Hughes1,Hughes2,Weiss}.


In the simplest design of a so called coined discrete-time quantum walk, a walker endowed with a two-dimensional internal coin state is transported along a line. Within each step the walker's internal state is first altered by a coin operator and subsequently a shift operator moves the walker in both directions consistently with the updated internal state. The walker's wave function spreads on the graph, repeatedly splits to reunite again giving rise to various interference patterns with corresponding non-classical behavior. While the classical walker undergoes a diffusive process, the evolution of its quantum counterpart is ballistic, resulting in quadratic speed up of walker's spreading through the lattice  \cite{quantum_random_walks}.

This simplest model has been quite soon after its introduction
\cite{review} generalized in a number of ways \cite{Mackay, Bach, Linden}.
There are higher dimensional coin walks and inhomogeneous walks (including coin point defects and different boundary conditions) which reveals another non-classical effect called trapping or localization \cite{inui:psa,inui:grover1,miyazaki,watabe,falkner,machida}. In such case, a part of the walker's wave function is captured in a vicinity of the origin and the efficiency of a walker's eventual transport between different regions can be, due to this effect, significantly reduced. A separate class of generalizations concerns the position space of the walker. Various types of the underlying graphs have been studied, e.g. quantum walks on general graphs \cite{Aharonov2001}, hypercubes \cite{Moore2002,Portugal2008}, trees \cite{Segawa2009}, honeycombs \cite{Lyu2015,Kendon2016}, spidernets \cite{Segava2013} or fractal structures \cite{Lara2013} (see also review \cite{review}). One of the main driving forces of this research activities is interest in asymptotic properties of quantum walks, including limiting position distributions, speed of walker's propagation, and structure of trapped states.

A different family of generalizations aims to incorporate and analyze the influence of imperfections and external perturbations on the behavior of quantum walks. While static spatial random changes of the coin may lead to Anderson localization \cite{yin:qw:loc,ahlbrecht:qw:loc}, temporal randomness in the coin operator typically causes decoherence resulting in a transition to classical behavior \cite{dec:brun,dec:kend}. External perturbations affecting the shift operator may correspond to randomly disappearing and again reappearing edges and are known as quantum walks on dynamically percolated graphs \cite{Kendon}. Percolation aspires to model and study, under certain conditions, transport \cite{Engel,Rebentrost} which is radically influenced by the connectivity of the inherent structure which carries the walker and is represented by a graph.

The present paper has several aims. First of all, the analysis of quantum walks on more complex graphs brings additional freedom, namely many equally appropriate possibilities how to choose the shift operator. Each shift operator generates a significantly different evolution. We supplement the list of quantum walks generalizations with one
capable to treat walks performed on arbitrary graphs in a unified manner and simultaneously allowing a classification of all the applicable shift operators for the walk. Second, the model is well designed to be further exploited in research of quantum walks on percolated graphs. Based on \cite{asymptotic2} we generalize an analytical method for the asymptotic regime of quantum walks on dynamically percolated finite graphs. Third, we apply the method to study Grover walk on dynamically percolated graphs with maximal degree three. We analyze two types of shift operators naturally defined on planar graphs. We show that the asymptotic evolution of the percolated walk with the reflecting (flip-flop) shift operator is characterized by a rich structure of trapped states. A general recipe which provides a basis of the trapped states is given.
Moreover, it is shown that also percolated walks with different shift operators exhibit the same structure of trapped states if we properly tune the coin operator. In contrast to this, the percolated Grower walk with cyclic shift operators has no trapped states. We present a simple criterion, based on a graph coloring, allowing to decide whether the asymptotic evolution of the percolated Grover walk with a given cyclic shift operator is stationary or not. Remarkably, if there is an edge-3-coloring of an associated state graph, we can define a cyclic shift operator for which  the percolated Grover walk has a non-stationary asymptotic evolution. Fourth, we discuss how the existence of trapped states reduces an overall efficiency of quantum transport and investigate the excitation transport modeled via the reflecting Grover walk on the percolated and unpercolated cube.

We briefly describe the structure of the paper. In section \ref{section2} we introduce a model for coined quantum walks on an arbitrary graph with an arbitrary shift operator, discuss its properties, and define coined quantum walks on dynamically percolated graphs. An analytical treatment of their asymptotic evolution is given in section \ref{section3}. In section \ref{section4} we discuss scenarios with restricted percolation. Section \ref{sec:3-regular graphs} is devoted to the analysis of the Grover walk on graphs with maximal degree three. A source-to-sink quantum transport with a simple example of the percolated and unpercolated cube is discussed in section \ref{section6}. We conclude in section \ref{section7}. Finally, in appendix \ref{app:graph theory} we provide basics of graph theory and in appendices \ref{app:asymptotics-proofs}, \ref{appendix_basis}, and \ref{appendix_non_p} detailed proofs of our claims are given.


\section{Coined Quantum Walk Definitions}
\label{section2}
In this section, we first recall the standard definition of a coined quantum walk and then we present our generalized definition. The new definition is demonstrated on a very simple but detailed example at the end of the section.

\subsection{Standard definition}

In a quantum walk, the walker is a quantum particle which is moving in
an environment associated with a graph, i.e. a set of vertices representing the
position of the walker and a set of edges representing his possible motion
from one position to the next. Due to his quantum nature, the
walker can be in a superposition of states (reside on several
vertices) and his paths can interfere.

The standard definition of a quantum walk starts with an undirected graph
$G(V,E)$ with the set of vertices $V$ and the set of
undirected edges $E$ typically representing some kind of lattice. The most common example is the quantum walk on a line (chain) and another can be a quantum walk on a square lattice \citep{asymptotic2}. The lattice has an associated Hilbert space
$\sH_p$, called the \textit{position space}. Therefore, the base vectors
in $\sH_p$ are $\ket{v}$ for $v\in V$. Further, the walker has internal degrees of freedom with a
corresponding Hilbert space $\sH_c$ called the \textit{coin space}. The name comes from the analogy with classical random walks, where the walker flips a coin to determine the direction of the next step.
States in the coin space correspond to directions in the lattice. For example "right" as $\ket{+}=
\left[
\begin{array}{ccr}
 1 \\
 0
\end{array}
\right]$ and "left" as $\ket{-}=
\left[
\begin{array}{ccr}
 0 \\
 1
\end{array}
\right]$ on the line graph. The whole quantum walk takes place on the Hilbert space $\sH=\sH_p
\otimes \sH_c$, where the state of the walker is given by a position in the lattice and a direction of further movement.

Coined quantum walks are discrete in time and every step is realized by a unitary evolution operator $U$ as
\begin{align}
\label{step}
\ket{\psi(t+1)}=U\ket{\psi(t)}=U^{t+1}\ket{\psi(0)}.
\end{align}
The evolution operator of a
quantum walk is a product of two unitary operators: $U=SC$, where
$C$ is called the \textit{coin operator} and $S$ is called the
\textit{shift operator}. The coin operator is an arbitrary unitary
operation acting on the coin space and, therefore, determining the
direction of the walker's movement at the given time step. The
actual movement is then realized by the shift operator. The walker
is displaced to the neighboring vertex according to the internal state, which is usually left unchanged.

Let us illustrate the dynamics on the walk on a line graph with the Hadamard coin
\begin{align}
\label{hadamard}
H &=
\frac{1}{\sqrt{2}}\left[
\begin{array}{ccr}
 1 & 1\\
 1 & -1
\end{array}
\right].
\end{align}
Initializing the system in the state $\ket{\psi(0)}=\ket{0,-}$, one step of the
walk then results in
\begin{align*}
\ket{\psi(1)}&=U\ket{\psi(0)}=S(I_p \otimes
H)\ket{0,-} = \\
&=S\frac{1}{\sqrt{2}}(\ket{0,+}-\ket{0,-}) = \\
&=\frac{1}{\sqrt{2}}(\ket{1,+} - \ket{-1,-}),
\end{align*}
where $I_p$ is the identity operator on the position space.
This definition of the shift operator relies heavily on a correspondence of global directions (here "left" and "right") with edges at all vertices. However for some graphs, like the honeycomb lattice or general irregular graphs, local directions of edges may vary from vertex to vertex. In this paper, we introduce a new definition, which coincides with the standard one on regular lattices and in a convenient way generalizes it in several directions. First, it extends the class of admissible graphs including non-regular graphs. Second, it allows for a simple classification of all
possible shift operators for a given graph and third, it can be easily accommodated for modeling external disturbances of the walk caused by randomly broken edges (percolation model).

\subsection{New definition - the Hilbert space}

The geometry of walker's possible positions and shifts is captured by the \textit{structure graph} $G(V,E)$ with the set of vertices $V$ and the set of edges $E$. Since the walker can move along every existing link in both directions, the structure graph is undirected. For the sake of simplicity of arguments, we further assume the structure graph to be connected and to have maximally countably infinite sets of vertices and edges. On the other hand, the structure graph is not required to be simple so both parallel edges and loops are allowed.

In the new definition, the directions of walkers shifts are understood locally and we identify
the \textit{base states} of the Hilbert space $\sH$ with directed edges of a
so called \textit{state graph}. The state graph $G^{(d)}(V,E^{(d)})$ has the same set of vertices
$V$ as the structure graph $G(V,E)$ and its set of directed edges
$E^{(d)}$ consists of two subsets: $E^{(d)} = E^{(d)}_p \cup
E^{(d)}_u$.

Edges from the first subset $E^{(d)}_p$ are called \textit{paired}
and are derived from the structure graph $G(V,E)$. For every
undirected edge $e \in E$ we have two directed edges $e^{(d)}_1,
e^{(d)}_2 \in E^{(d)}_p$ oriented in opposite directions and
connecting the same two vertices as~$e$. Corresponding to these
paired edges we introduce \textit{paired states} $\ket{e^{(d)}_1},
\ket{e^{(d)}_2}$. Edges in the second subset $E^{(d)}_u$ are called \textit{unpaired}
and are independent of the structure graph $G(V,E)$. Unpaired edges
are loops $e^{(d)} \in E^{(d)}_u$ with their corresponding states $\ket{e^{(d)}}$. Note that caution is needed when working with non-simple graphs, where the presence of paired loops is also possible. Adding these \textit{unpaired loops} allows us to arbitrarily increase the degree of chosen vertices. It
may be for example on borders of finite graphs \citep{asymptotic2}
or if there is some state representing no movement of the walker
\citep{three_state}.

Overall, for every directed edge $e^{(d)} \in E^{(d)}$ going from
$v_1\in V$ to $v_2\in V$ (possibly $v_1=v_2$), there is an associated base state
$\ket{e^{(d)}}$ and the Hilbert space is the span of all these states. The state $\ket{e^{(d)}}$ represents a
walker standing at vertex $v_1$ and having the direction towards $v_2$.
An example of a structure graph with its associated state graph is provided in Fig. \ref{fig:simple_example}.

For a general graph the Hilbert space $\sH$ of a quantum walk does not have to be of the tensor product form
$\sH=\sH_p \otimes \sH_c$, but it can always be written as a direct sum
of vertex subspaces: $\sH=\bigoplus_{v\in V}\sH_{v}$. A \textit{vertex subspace} $\sH_{v}$ is spanned by states corresponding to edges originating in $v\in V$.

\subsection{New definition - the time evolution}
\label{time_evolution}

The time evolution proceeds in discrete time steps and is governed
by a unitary evolution operator $U$ according to the equation
(\ref{step}), where the action of $U$ can be split into the coin operation $C$ and the shift operation $S$. The role of the coin operator $C$ remains the same - it mixes states in vertex subspaces. The difference is that
the dimensions of these vertex subspaces $\sH_v$ may vary for
various vertices $v\in V$. Therefore, the coin operator is in
general no longer of the form $C=I_p \otimes C_0$, but still can be
expressed as $C=\bigoplus_{v\in V}C_v$. We also allow different
coins even at vertices of the same degree.

The shift operator $S$ moves the walker among vertices. In the case of a quantum walk on a line graph,
there is a natural shift operator: the walker keeps moving
in one direction (while also being influenced by the coin operator in every step). Nevertheless, for more complex graphs, e.g. for the honeycomb lattice, the choice of the shift operator may be far from obvious. Typically, there are multiple possibilities
without any of them being naturally preferred. Since the choice of
the shift operator is crucial for the resulting time evolution of the quantum
walk, we need a general framework capable of covering all possible shift
operators and allowing for the analysis of the walker's behavior.

Our starting point is a canonical shift operator available for any graph. We denote this particular shift operator by
$R$ and refer to it as the \textit{reflecting shift operator} (the
name flip-flop shift operator is sometimes used in the literature \citep{flipflop}). The action of $R$ is defined as follows. If
we have an undirected edge $e\in E$ with two corresponding directed
paired edges $e^{(d)}_1,e^{(d)}_2 \in E^{(d)}_p$, then
$R\ket{e^{(d)}_1}=\ket{e^{(d)}_2}$ and
$R\ket{e^{(d)}_2}=\ket{e^{(d)}_1}$. Any unpaired state $\ket{e^{(d)}_l}$
for $e^{(d)}_l \in E^{(d)}_u$ is mapped to itself, so
$R\ket{e^{(d)}_l}=\ket{e^{(d)}_l}$. We note that
the operator $R$ is its own inverse and since it is unitary, it is
also Hermitian ($R^{-1}=R=R^\dagger$). An example of the action of the reflecting shift operator is given in
Fig. \ref{fig:simple_example_reflecting}.

Equipped with the reflecting shift operator we can design any possible shift operator in a convenient way. It is achieved in two stages. First we apply the reflecting shift operator and then we change the walker's direction encoded into its coin state. In more details, the reflecting shift operator moves the walker in direction of the walker's current coin state and sets the coin state to the one associated with the reverse directed edge. Afterwards, we change walker's direction by a subsequent application of a \textit{local permutation} operator $P$ acting locally on each vertex subspace $\sH_v$. Thus in the next step, the walker, instead of going back, will follow a new desired direction. Consequently, any shift operator $S$ consist of an application of the reflecting shift operator followed by a particular local permutation, i.e. $S=PR$.

\begin{figure}
    \centering
    \includegraphics[width=190 pt]{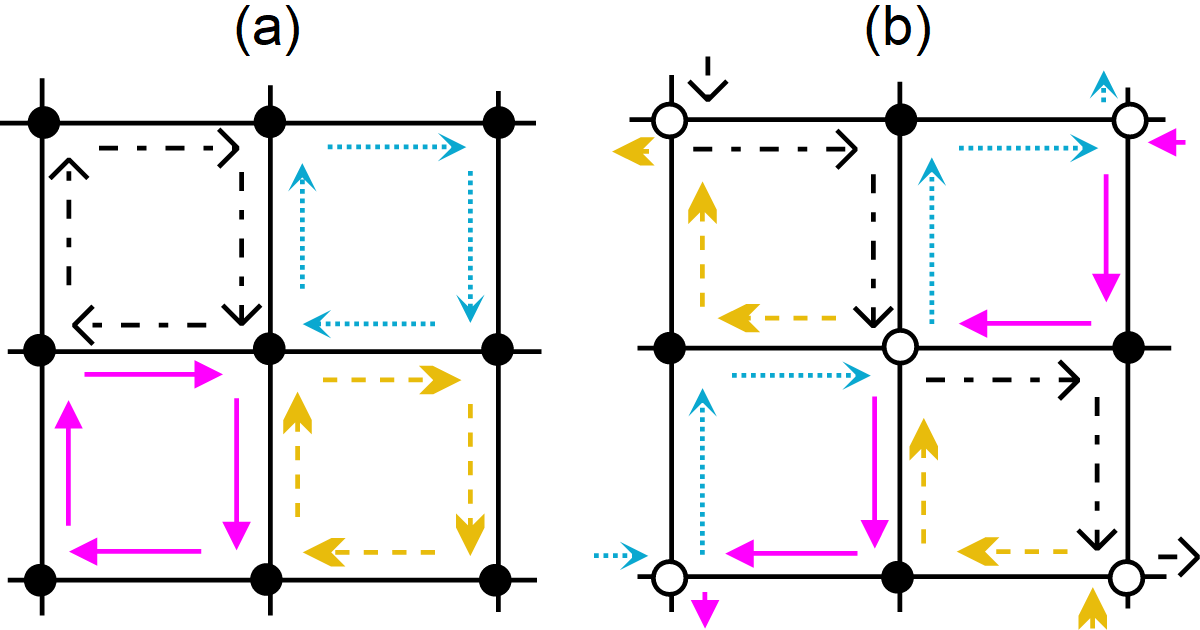}
    \caption{(color online) Actions of shift operators depicted by colors and line types for operators with (a) counter-clock-wise rotations applied as local permutations in all vertices resulting in a cyclic movement and (b) clock-wise (white vertices) and counter-clock-wise (black vertices) rotations distributed evenly giving a zig-zag diagonal motion. These are alternatives to the standard shift operator used on a square lattice, where the walker always keeps his direction of movement.}
    \label{fig:different_shifts_square}
\end{figure}

Due to the locality, $P$ can be written in a block diagonal form $P=\bigoplus_{v\in V}
P_v$, where $P_v$ is the local permutation in the vertex $v\in V$.
Every operator $P$ determines one possible shift operator $S=PR$.
Therefore, there are in principle $\Pi_{v\in V}(\mathrm{d}(v)!)$ different shift operators, however the action of $P$ is typically chosen to be the same
at all vertices or in significant fractions of vertices. A simple example is shown in Fig. \ref{fig:simple_example_transporting} and a more complex one is presented in Fig. \ref{fig:different_shifts_square}.

%

The evolution operator for one step of a quantum walk is frequently defined as
\begin{align*}
U^{(1)}=SC=PRC,
\end{align*}
where the index (1) refers to the position of the coin
operation $C$ in $U$. The fact that we first apply the coin operator
and then the shift operator comes from the analogy with a classical
random walk. The walker flips a coin and makes a step according to
the result of the coin flip. Nevertheless, in quantum walks there is no principal need for applying a shift and a coin operator in the order $SC=PRC$. In particular, we extensively use the variant $U^{(3)}=CPR$ in this work. The reason is threefold.

First, and most importantly, the variant $U^{(3)}=CPR$ is very convenient for the investigation of the asymptotic evolution of dynamically percolated quantum walks. It allows us to obtain results in a very simple and elegant form - in particular a so called shift condition, which is described later.

Second, there is a simple relationship between evolutions generated by $U^{(1)}$ and $U^{(3)}$. Indeed, starting in the state $\ket{\psi(0)}$, the state after
$n$ steps of a walk with $U^{(3)}$ is $\ket{\psi^{(3)}(n)} = \left(
U^{(3)} \right)^n \ket{\psi(0)} =  C \left( U^{(1)} \right)^n
C^\dagger \ket{\psi(0)}$. Therefore, the difference in the final
state is equivalent to changing the initial state of the walk and
applying one final unitary operation, which is local at vertices. We
will later use similar relation to obtain the asymptotic behavior of a percolated quantum walk generated by the operator $U^{(1)}$ using the
solution for $U^{(3)}$, which is easier to find.

Third, the variant $U^{(3)}$ reveals a close relationship between the choice of the coin operator and the shift operator - a different local
permutation $P$ can also be achieved by modifying the coin. In our
formalism, we immediately see that $U^{(3)}=C_1 S = C_1 (P R) = (C_1
P) R \equiv C_2 R$, i.e. the walk with an arbitrary shift
operator $S$ and the coin $C_1$ can also be viewed as a walk
with the reflecting shift operator $R$ and the coin $C_2 = C_1 P$.
Therefore, all claims about properties of
quantum walks with a particular coin operator should be accompanied by
a discussion of the choice of the shift operator. The existence of modified coins and shift operators that together
result in the same evolution can be found effortlessly in our
formalism, but it would be difficult in the standard definition. This can be considered as an advantage of the new formalism. Note also, that there is no reason for considering $U^{(2)}=PCR$ as a separate variant, since it is just the variant $U^{(3)}$ with a different coin.


Overall, our formalism of local permutations allows for a
convenient classification of possible choices of shift operators and
gives them an intuitive interpretation.
This is advantageous for complex graphs without a privileged shift operator
available, but also for simple regular graphs, where alternative
options can be investigated. We also show that different local permutations and therefore different shift
operators are basically equivalent to choosing different coins. Moreover, the formalism is also well suited for the investigation of walks with dynamical percolation, which is presented in detail in the following text.

\subsection{Summary of the new definition}

A quantum walk is fully determined by four choices. First, there is
the underlying structure graph $G(V,E)$ extended to the associated
directed state graph $G^{(d)}(V,E^{(d)})$ by replacing every
undirected edge by two directed edges (corresponding to paired
states) and by adding unpaired loops (corresponding to unpaired
states). The structure graph then defines the reflecting shift
operator $R$. The second choice is the coin operator $C$. In
principle, the coin can act in every vertex as an arbitrary unitary
operator of the dimension equal to the degree of this vertex. The third choice is the shift operator $S$
or equivalently the local permutation $P$. For the application of $C$ and $P$ as matrices, ordering of states in vertex subspaces must be fixed. The last choice is the order of operators $S=PR$ and $C$ in $U$.

One might want to define a quantum walk on an arbitrary directed graph. In our framework this means to introduce directed unpaired edges connecting two distinct vertices and its associated base states. In a general case the
reflecting shift operator $R$ is not available anymore. However, if the
in-degree and the out-degree in every vertex of the graph is the same, one can always start with some shift operator given by an Eulerian cycle and then
use the formalism of local permutations to classify the remaining
shift operators.

\subsection{Percolated quantum walk}
\label{sec:Dynamically percolated quantum walk}
Originally, percolation is a concept from graph theory unrelated to
quantum physics. Starting from some regular lattice each edge is made open with a chosen and fixed probability $p$ and closed with the probability $1-p$. We ask whether there is an infinite component of continuity. The formation of such a component has the character of a phase transition at a critical probability $p_c$.
If $p\geq p_c$, an infinite cluster of open edges is present in the resulting percolation graph with probability 1 \citep{percolation}. The appearance of the cluster rapidly changes the global properties of the system and is hence of significant interest in physics.

In the context of quantum walks, the term \textit{percolation} refers to random disturbances of a quantum walk by closing some edges as described above and we call the evolution on such graphs a \textit{percolated quantum walk}. A single realization of the percolation process gives rise to a set of open edges (a configuration) $K\subset E$, for which each paired edge in the state graph additionally becomes either open or closed. As the walker cannot pass closed edges, naturally a modification of the reflecting shift operator is needed. The modified reflecting operator $R_K$ treats the closed paired edges as unpaired loops. To provide its rigorous description we introduce the following notation. If $i\in E_p^{(d)}$ is a paired edge, we denote its counterpart edge in the pair $\tilde{i}$ and for an unpaired edge $i\in E_u^{(d)}$ let $\tilde{i}=i$. Thus $\tilde{\tilde{i}}=i$ is valid for any edge. We stress that this notation refers to the state graph and is independent of a particular choice of configuration $K$ in the percolated quantum walk. Using this notation the original reflecting operator $R$ takes the form
\begin{align}
\label{shift_operator_perc}
R = \sum_{i\in E^{(d)}}\ket{\tilde{i}}\bra{i},
\end{align}
and the modified reflecting operator $R_K$ for a given configuration $K$ can be described as
\begin{align}
\label{shift_operator_perc}
R_K = \sum_{i\in E^{(d)}}\ket{k(i)}\bra{i},
\end{align}
where the permutation map $k$ is defined as $k(i)=\tilde{i}$ for an open paired edge, $k(i)=i$ for a closed paired edge, and $k(i)=i=\tilde{i}$ for an unpaired edge. Note that also for the percolated walk $R_K = R_K^{-1}=R_K^\dagger.$

The coin operation $C$ is not altered by the percolation and also
the local permutation $P$ stays the same independently of the
configuration $K$. Therefore, the new evolution operator $U_K$ is only given by the modified reflecting shift operator $R_K$. We may note that the combined action of $R_K$ and
$P$ for a closed edge is such that the walker stays at the vertex
and he is mapped by $P$ to the state which would otherwise be the
end state of the walker coming from the opposite direction.

\subsection{Simple example}

In this part we present a simple example of a quantum walk described using our developed framework. This example serves for demonstration of the concepts presented above.

\begin{figure}
    \centering
    \includegraphics[width=130 pt]{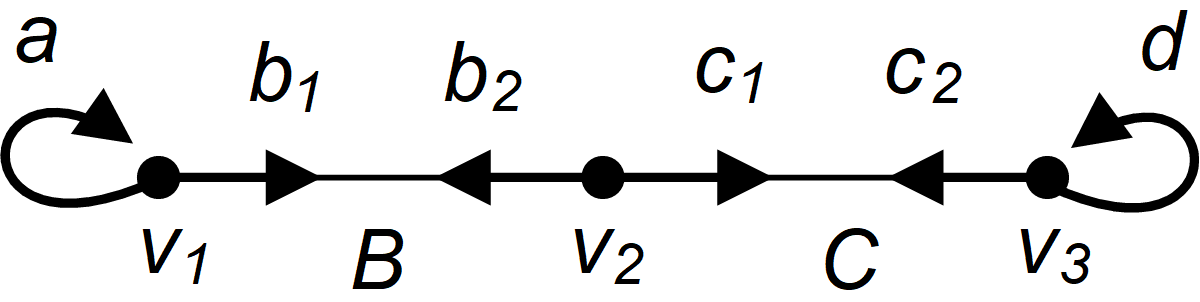}
    \caption{An example of the structure graph and its associated state graph. Both graphs share three vertices $V=\{v_1,v_2,v_3\}$. The structure graph has two undirected edges $E=\{B,C\}$ and the state graph has 6 directed edges $E^{(d)}=\{a,b_1,b_2,c_1,c_2,d\}$.}
    \label{fig:simple_example}
\end{figure}

Both the structure graph and the associated state graph of the walk are depicted in Fig. \ref{fig:simple_example}. The structure graph $G(V,E)$ has only three vertices $V=\{v_1,v_2,v_3\}$ and two undirected edges $E=\{B,C\}$. The associated state graph $G^{(d)}(V,E^{(d)})$ has the same set of vertices $V$, but the set of (directed) edges is $E^{(d)}=\{a,b_1,b_2,c_1,c_2,d\}$. Four of these directed edges are derived from the structure graph and therefore are denoted as paired: $E^{(d)}_p=\{b_1,b_2,c_1,c_2\}$. The remaining directed edges $E^{(d)}_u=\{a,d\}$ are added loops and therefore are unpaired. The Hilbert space of the quantum walk is $\sH=\mathrm{span}\{\ket{a},\ket{b_1},\ket{b_2},\ket{c_1},\ket{c_2},\ket{d}\}$ with vertex subspaces $\sH_{v_1}=\mathrm{span}\{\ket{a},\ket{b_1}\}$, $\sH_{v_2}=\mathrm{span}\{\ket{b_2},\ket{c_1}\}$ and $\sH_{v_3}=\mathrm{span}\{\ket{c_2},\ket{d}\}$.

\begin{figure}
    \centering
    \includegraphics[width=130 pt]{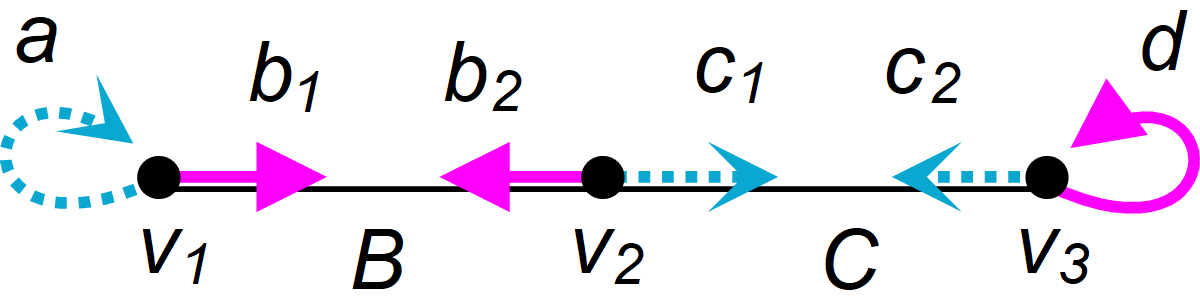}
    \caption{(color online) The action of the reflecting shift operator on the line graph with three vertices. The operator moves the walker along directed edges and the final coin state is indicated by colors and line types. As a result, unpaired loops are mapped to themselves.}
    \label{fig:simple_example_reflecting}
\end{figure}

The action of the reflecting shift operator $R$ on our graph is shown in Fig. \ref{fig:simple_example_reflecting}. For paired states, we have $R\ket{b_1}=\ket{b_2},R\ket{b_2}=\ket{b_1},R\ket{c_1}=\ket{c_2},R\ket{c_2}=\ket{c_1}$. For unpaired loop states, the action is $R\ket{a}=\ket{a}$ and $R\ket{d}=\ket{d}$.

\begin{figure}
    \centering
    \includegraphics[width=130 pt]{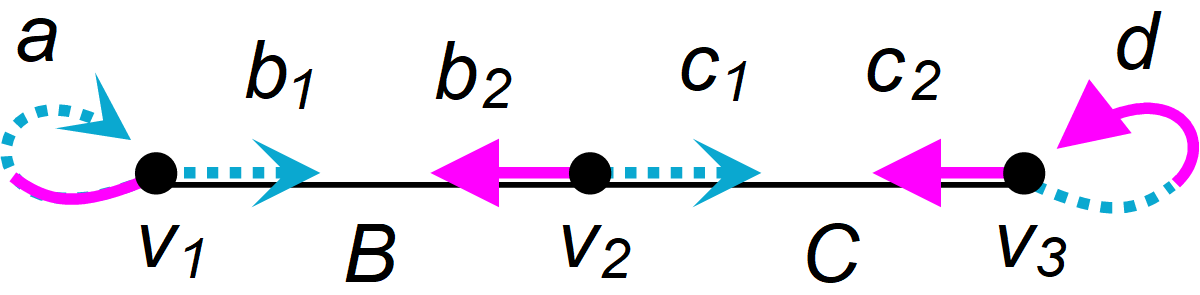}
    \caption{(color online) The standard shift operator on the line graph with three vertices.  As indicated, it moves the walker along directed edges mapping each arrow-head to arrow-tail of the same color and line type.}
    \label{fig:simple_example_transporting}
\end{figure}

The reflecting shift operator is not the shift operator usually used for a quantum walk on a line. The shift operator is typically chosen to keep the walker in the original direction. In our formalism, this is achieved by using the swap operator $\sigma_x$ as the local permutation at each vertex $v$. The corresponding shift operator has the form $S=PR=\left(\bigoplus_{v\in V} \sigma_{x}\right) R$. Explicitly, the action of $S$ is $S\ket{a}=\ket{b_1}, S\ket{b_1}=\ket{c_1}, S\ket{c_1}=\ket{d}, S\ket{d}=\ket{c_2}, S\ket{c_2}=\ket{b_2}, S\ket{b_2}=\ket{a}$ as can be seen in Fig. \ref{fig:simple_example_transporting}.

The identity $I_2$ and $\sigma_x$ are the only two possible permutations on a two-dimensional vertex subspaces. Therefore, all possible shift operators for walks on line graphs can be obtained by distributing local permutations $I_2$ and $\sigma_x$ among vertices.

\begin{figure},
    \centering
    \includegraphics[width=150 pt]{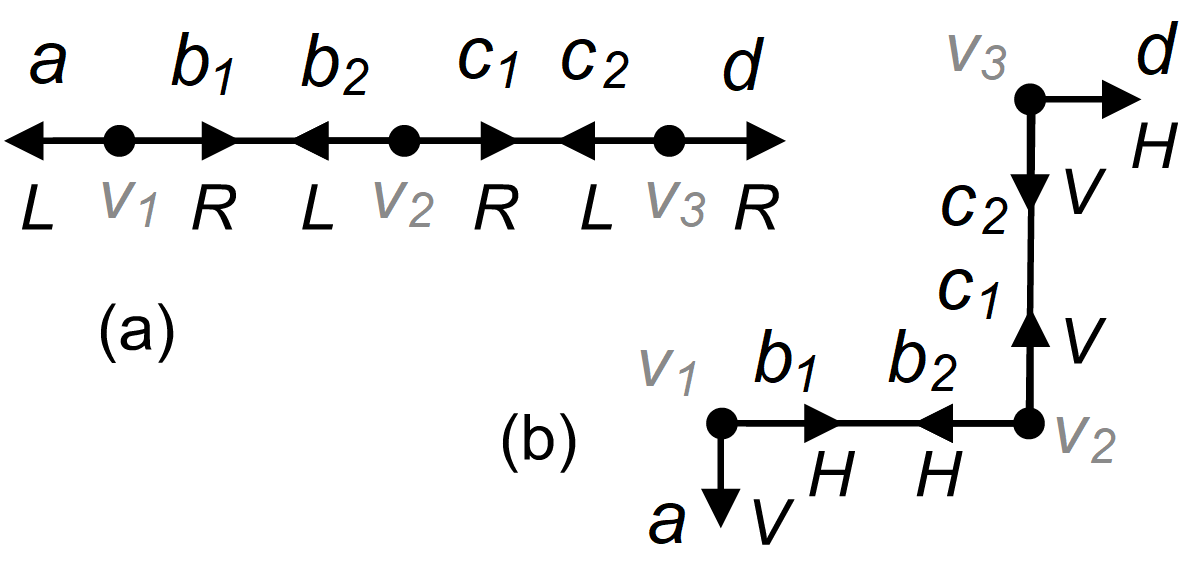}
    \caption{The graph presented in Fig. \ref{fig:simple_example} with two different directional labelings of edges: (a) left/right and (b) vertical/horizontal.}
    \label{fig:simple_example_labeling}
\end{figure}

Let us illustrate the importance of the chosen order of base states. Assume, for example, the Hadamard coin (\ref{hadamard}). This matrix is asymmetrical in the sense that it adds additional phase -1 when it is applied to the second state of the basis. For this reason, the way how we order base states in each vertex subspace is relevant. For example, if we use the one shown in Fig. \ref{fig:simple_example_labeling} (a) and order the states as $(\ket{L},\ket{R})$ at all vertices, in terms of directed edges we have $(\ket{a},\ket{b_1})$, $(\ket{b_2},\ket{c_1})$ and $(\ket{c_2},\ket{d})$. If, on the other hand, our graph is for example placed in a square lattice, we follow Fig. \ref{fig:simple_example_labeling} (b) and order the states as $(\ket{V},\ket{H})$ at all vertices, in terms of directed edges we have $(\ket{a},\ket{b_1})$, $(\ket{c_1},\ket{b_2})$ and $(\ket{c_2},\ket{d})$. If we do not use the modified coin
\begin{align}
H' &= \sigma_x H \sigma_x^\dagger =
\frac{1}{\sqrt{2}}\left[
\begin{array}{ccr}
 -1 & 1\\
 1 & 1
\end{array}
\right]
\end{align}
in the vertex $v_2$, the dynamics will differ compared to the dynamics generated by the Hadamard coin on the line graph \ref{fig:simple_example_labeling} (a).

\begin{figure}
    \centering
    \includegraphics[width=120 pt]{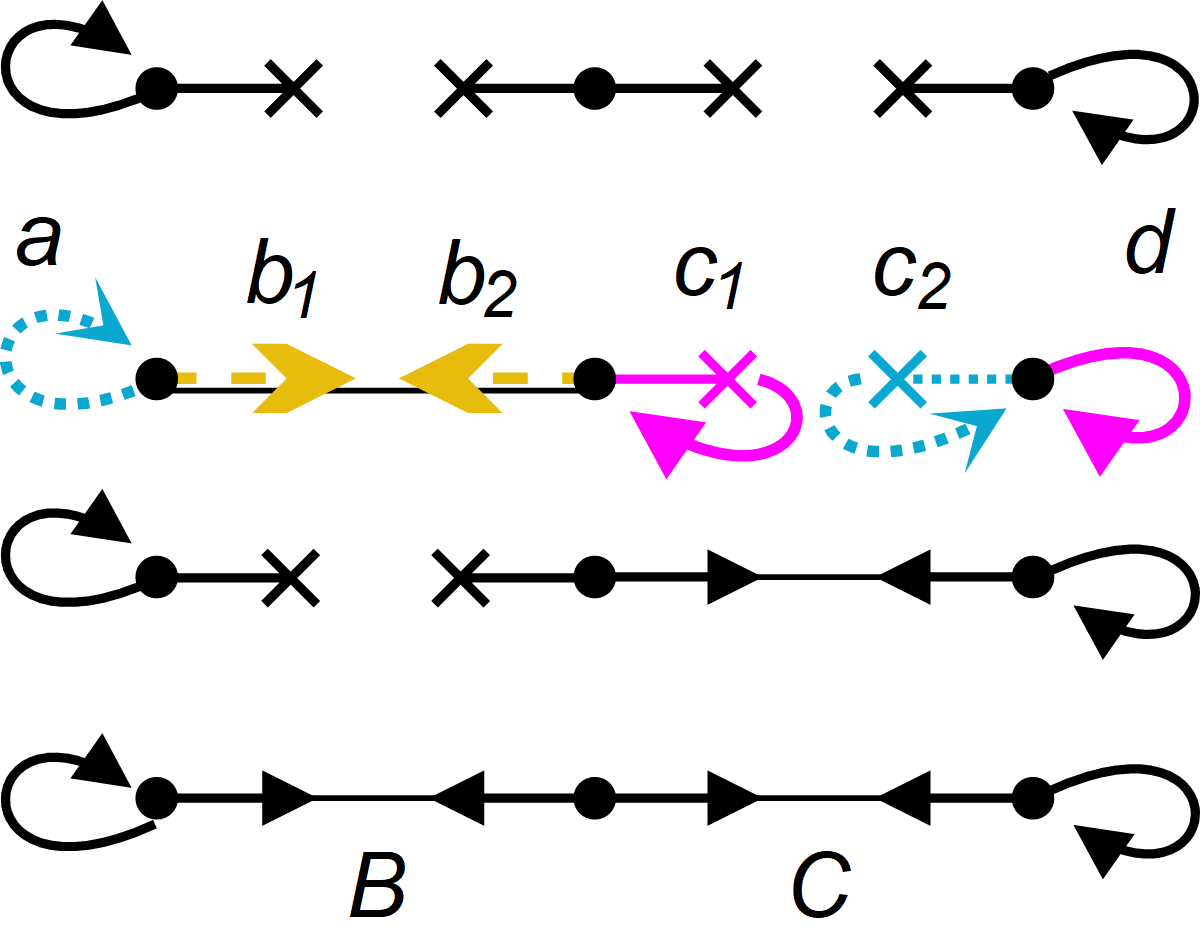}
    \caption{(color online) All four percolation configurations of the graph presented in Fig. \ref{fig:simple_example}. Broken paired edges are represented by crosses at arrow points. The second configuration $K_B$ is accompanied with the action of the modified reflecting shift operator $R_B$. The operator moves the walker along directed edges except the case of broken paired edges - those are treated as unpaired loops and the walker stays in the original vertex. The final coin state is indicated by colors and line types.}
    \label{fig:simple_example_percolation}
\end{figure}

When we introduce percolation by making every edge present with a probability $p$ and missing with the probability $1-p$, we have a set of possible configurations of the percolation graph $2^E= \{\emptyset, \{B\}, \{C\}, \{B,C\}\}\equiv\{K_\emptyset, K_B, K_C, K_{BC}= K_E\} $ with probabilities of occurrence $\pi_{K_\emptyset}=(1-p)^2$, $\pi_{K_B}=\pi_{K_C}=p(1-p)$ and $\pi_{K_E}=p^2$. For example, for the configuration $K_B$ the modified reflecting operator $R_B$ acts as: $R_B\ket{a}=\ket{a}, R_B\ket{b_1}=\ket{b_2}, R_B\ket{b_2}=\ket{b_1}, R_B\ket{c_1}=\ket{c_1}, R_B\ket{c_2}=\ket{c_2}, R_B\ket{d}=\ket{d}$ as shown in Fig. \ref{fig:simple_example_percolation}. Therefore, the walker cannot pass the edge $C$ and the edges $c_1$ and $c_2$ are referred to as broken paired edges.

\section{Asymptotic Evolution of Dynamically Percolated Quantum Walks}
\label{section3}

In this part we generalize the concept of dynamically percolated quantum walks originally introduced for finite $1D$ and $2D$ lattices \cite{asymptotic1,asymptotic2} and present a universal procedure allowing for a solution of their asymptotic dynamics. From now on we assume a finite structure graph $G(V,E)$ with a finite number of vertices $\#V$ and a finite number of edges $\# E$.

The dynamically percolated quantum walk captures the situation in which the walker is moving on a graph which may randomly change in each step. Edges closed in one step can be reopened in further steps and vice versa.
A new configuration of open edges $K$ is randomly generated with probability $\pi_K$ for each step of the walker's evolution, which is subsequently given by the corresponding unitary operator $U_K$. However, as the actual configuration $K$ is unknown, the dynamical percolation introduces classical uncertainty to the overall walker's evolution. Taking into account all possible configurations, one step of the dynamically percolated quantum walk given by the super-operator $\mathcal{S}$ maps walker's state $\rho(t)$ at time $t$ onto the state
\begin{equation}
\rho(t+1)= \mathcal{S}\left(\rho(t)\right)=\sum_{K\subset E}\pi_KU_K \rho(t) U_K^\dagger.
\label{timeevolutionpercolated}
\end{equation}
The evolution (\ref{timeevolutionpercolated}) is governed by a random unitary operation $\mathcal{S}$, which is a certain subclass of open system's dynamics also often called an external field \cite{Alicki1987}. In general an analytical treatment of such evolutions is hard, but it can be significantly simplified if we are interested in their asymptotic behavior.

A procedure for determining the asymptotic behavior of a system governed by random unitary operations was suggested in \cite{ruo}. The asymptotic dynamics is determined by so called \textit{attractors} -- eigen-matrices of the generator $\mathcal{S}$ associated with eigenvalues from an \textit{asymptotic spectrum} (eigenvalues of the map $\mathcal{S}$ with modulus one). Both the attractors and the asymptotic spectrum are solutions of attractor equations. In particular, an operator $X_{\lambda}$ is an attractor corresponding to $\lambda$ if it satisfies  attractor equations
\begin{equation}
U_K X_\lambda U_K^\dagger = \lambda X_\lambda\,, ~~{\rm for~all}~K\in 2^E
\label{attractors}
\end{equation}
and $|\lambda|=1$ ($\lambda$ is from the asymptotic spectrum). Note that the actual values of nonzero probabilities $\pi_K$ do not affect the asymptotic dynamics. Provided we have found an orthonormal basis $\left\{X_{\lambda,i}\right\}$ of attractors with respect to the Hilbert-Schmidt scalar product, i.e. $\Tr (X_{\lambda_1,i}^{\dagger}X_{\lambda_2,j})=\delta_{\lambda_1 \lambda_2} \delta_{ij}$, the asymptotic dynamics (the limit for infinitely many steps) of dynamically percolated quantum walk is given as \cite{ruo}
\begin{equation}
\label{as_state}
\rho_{t \rightarrow \infty}(t) = \sum_{\lambda,i}\lambda^t\Tr{\left(\rho(0) X_{\lambda,i}^\dagger \right)} X_{\lambda,i}.
\end{equation}
Index $i$ distinguishes different attractors for a given eigenvalue
$\lambda$ and $\rho(0)$ is the initial state of the quantum
walk.

In order to obtain attractors we now apply the procedure described in \cite{asymptotic1} employing the special structure of unitary operators $U_K$. We formulate the whole approach for the variant $U_K^{(3)}=CS_K=CPR_K$ of the evolution operator.

The set of equations (\ref{attractors}) can be rewritten as
\begin{align}
\label{separated}
R_K X R_K^\dagger &= \lambda (CP)^\dagger X (CP)\,, ~~{\rm for~all}~K\subset 2^E.
\end{align}
The right-hand side is independent of the actual configuration $K$. It suggests to solve (\ref{separated}) in two subsequent steps. First, we choose the configuration with all edges closed, i.e. $K=\emptyset$, for which the modified reflecting shift operator simplifies to $R_\emptyset = I$, where $I$ is the identity operator. The equation (\ref{separated}) turns into the so-called \textit{coin condition}, which reads
\begin{align}
CP X (CP)^\dagger &= \lambda X.
\label{one}
\end{align}
As neither of the operators $C$ and $P$ mixes states from different vertex subspaces $\sH_v$, the matrix $CP$ is block-diagonal with respect to a properly chosen basis from these subspaces. Thus we can split the attractor matrix $X$ into blocks $X^{v_1}_{v_2}$ corresponding to pairs of vertices $v_1,v_2 \in V$ and solve the coin condition (\ref{one}) locally (for each attractor block $X^{v_1}_{v_2}$ individually)
\begin{equation}
\label{locally}
(C_{v_1} P_{v_1})X^{v_1}_{v_2}(C_{v_2} P_{v_2})^\dagger = \lambda X^{v_1}_{v_2},
\end{equation}
where $C_{v_1},C_{v_2}$ and $P_{v_1},P_{v_2}$ are blocks of operators $C$ and $P$ respectively acting on subspaces $\sH_{v_1}, \sH_{v_2}$ for vertices $v_1,v_2\in V$. By rearranging columns of the attractor matrix block $X^{v_1}_{v_2}$ into a vector $x^{v_1}_{v_2}$ (defined by $\braket{a,b|x^{v_1}_{v_2}}=\braket{a|X^{v_1}_{v_2}|b}$ for all $\ket{a}\in\sH_{v_1},\ket{b}\in\sH_{v_2}$), the equation (\ref{locally}) turns into a standard eigenvalue problem
\begin{equation*}
(C_{v_1} P_{v_1})\otimes(C_{v_2} P_{v_2})^\ast x^{v_1}_{v_2} = \lambda x^{v_1}_{v_2},
\end{equation*}
where the asterisk denotes complex conjugation. We find out that the spectral decompositions of local operators $C_vP_v$ determine the asymptotic spectrum and matrix blocks $X^{v_1}_{v_2}$ of any attractor. However, these blocks are further restricted by conditions emerging from the following second step of the procedure.

Since the right-hand side of the set of equations (\ref{separated}) is the same for all configurations, the left-hand sides must be equal for all configurations
\begin{equation}
\label{shift_cond}
R_K X R_K^\dagger = R_L X R_L^\dagger\,, ~~{\rm for~all}~K,L\in 2^E.
\end{equation}
We call (\ref{shift_cond}) the \textit{shift condition} and it tells how individual attractor blocks $X^{v_i}_{v_j}$ corresponding to different
pairs of vertices are bound together into one attractor matrix $X$. The shift condition is investigated in detail in a separate section \ref{sec_The Shift_Condition} below.

Let us discuss the relation between the asymptotic evolution of dynamically percolated quantum walks for the two variants of the unitary evolution operator $U_K$. The attractor equations (\ref{attractors}) for the variant with $U^{(1)}_K=PR_K C$ can be rewritten as $CPR_K CXC^\dagger R_K^\dagger P^\dagger C^\dagger = \lambda CXC^\dagger$. We realize that each attractor $X$ of the variant $U^{(1)}$ is in one-to-one correspondence with an attractor $W=C X C^\dagger$ for the variant $U^{(3)}$. Further, thanks to (\ref{one}) the transformation can be performed using $P$ instead of $C$. Therefore, we have revealed a simple one-to-one correspondence between attractors of both variants $U^{(1)}$ and $U^{(3)}$ given only by local permutations in vertex subspaces. Note in particular that for the reflecting walk (where $P=I$) the attractors are the same for both variants.

\subsection{p-Attractors}
\label{sec:p-Attractors}
In this part we recall a method further simplifying the search for attractors \cite{asymptotic2}. A significant part of attractors can be constructed from \textit{common eigenstates} of all unitary operators $U_K$, i.e. states following for some $\alpha$ the set of equations
\begin{equation}
\label{p-attractors}
U_K \ket{\phi_{\alpha}} = \alpha \ket{\phi_{\alpha}}\,, ~~{\rm for~all}~K\subset 2^E.
\end{equation}
Provided we are equipped with a basis $\ket{\phi_{\alpha, i}}$ of all common eigenstates ($i$ distinguishes different common eigenstates corresponding to a given $\alpha$), any linear combination
\begin{equation}
\label{p_attractors_construction}
Y_\lambda = \sum_{\alpha\beta^* = \lambda} A^{\alpha, i}_{\beta, j} \ket{\phi_{\alpha, i}}\bra{\phi_{\beta, j}}
\end{equation}
constitutes an attractor corresponding to the eigenvalue $\lambda = \alpha\beta^*$. Note that the unitarity of operators $U_K$ implies $|\alpha|=|\beta|=|\lambda|=1$.
Attractors of this type are called \textit{p-attractors} and they coincide with solutions of a set of equations
\begin{align}
\label{p_Y}
U_{K_1} Y_{\lambda, i} U_{K_2}^\dagger = \lambda Y_{\lambda, i}\,, ~~{\rm for~all}~{K_1},{K_2}\subset 2^E,
\end{align}
where the operators $U_{K_1}$ and $U_{K_2}$ can be different \cite{asymptotic2}. This set contains the same coin condition for the empty configuration (\ref{one}), but the shift condition for p-attractors is more restrictive and takes the form
\begin{equation}
\label{shift_p_attractors}
R_{K_1} X R_{K_2}^\dagger = R_{L_1} X R_{L_2}^\dagger\,, ~~{\rm for~all}~K_1,K_2,L_1,L_2 \in 2^E.
\end{equation}
Apparently, not any attractor is a p-attractor and to complete the attractor space we have to add additional elements - the \textit{non-p-attractors}. However, in this paper, we study percolated Grower walk on graphs with maximal degree three for two different types of shift operators and we show that the only needed non-p-attractor is the identity operator (see appendix \ref{appendix_non_p}). This considerably simplifies the subsequent analysis of walker's asymptotic behavior on these percolated graphs.

Indeed, to obtain common eigenstates requires significantly less effort. The set of equations (\ref{p-attractors}) for $U^{(3)}=CPR$ reads
\begin{align*}
R_K \ket{\phi_{\alpha, i}} &= \alpha (CP)^\dagger \ket{\phi_{\alpha, i}}  ~~{\rm for~all}~K\subset 2^E,
\end{align*}
with only the left-hand side being dependent on $K$. Ana\-lo\-gous\-ly as in the previous section, we use the empty configuration $K = \emptyset$ resulting in the coin condition for common eigenstates
\begin{align}
\label{p_one}
C P \ket{\phi_{\alpha, i}} &= \alpha \ket{\phi_{\alpha, i}}.
\end{align}
It is an eigenvalue equation, which can be readily solved in each subspace $\sH_v$ separately. It determines possible eigenvalues $\alpha$ and the form of associated common eigenvectors in each subspace $\sH_v$.

Employing all other configurations $K$ we arrive at the shift condition for common eigenstates, which binds together parts of a common eigenstate corresponding to different vertex subspaces $\sH_v$
\begin{align}
\label{p_shift_cond}
R_K \ket{\phi_{\alpha, i}} = R_L \ket{\phi_{\alpha, i}} \,, ~~{\rm for~all}~K,L\in 2^E.
\end{align}
The details how shift conditions affect the explicit form of common eigenstates is left for the next part \ref{sec_The Shift_Condition}.

Finally let us note that a similar procedure may be worked out for common eigenstates of dynamically percolated quantum walks with the variant $U^{(1)}=PRC$. In particular, one can show that common eigenstates for the variant $U^{(1)}=PRC$ are obtained from solutions for $U^{(3)}$ as $C^{\dagger} \ket{\phi_{\alpha, i}}$, which also equals $P \ket{\phi_{\alpha, i}}$.

\subsection{The shift condition}
\label{sec_The Shift_Condition}
The difference between attractors and p-attractors is only given by their shift conditions (\ref{shift_cond}) and (\ref{shift_p_attractors}), which they must obey. Let us explore and compare these conditions for attractors, p-attractors and common eigenstates. Assuming a general vector $\ket{\phi}=\sum_{j\in E^{(d)}}\phi_j\ket{j}$  and using (\ref{shift_operator_perc}) with $R_K = R_K^\dagger$, the shift conditions for common eigenstates
(\ref{p_shift_cond}) may be rewritten as
\begin{align*}
\sum_{i\in E^{(d)}}\phi_{k(i)} \ket{i} = \sum_{i\in E^{(d)}}\phi_{l(i)} \ket{i}\,, ~~{\rm for~all}~K,L\in 2^E,
\end{align*}
which turns into equality of vector elements
\begin{align*}
\phi_{k(i)} = \phi_{l(i)}\,, ~~{\rm for~all}~i\in E^{(d)},~K,L\in 2^E.
\end{align*}

Finally, each paired edge is open for some configurations and closed for the others. Therefore, the shift condition (\ref{p_shift_cond}) reduces to a simple rule
\begin{align}
\label{p_shift}
\phi_{i}=\phi_{\tilde{i}}\,, ~~{\rm for~all}~i\in E^{(d)}.
\end{align}
Note that (\ref{p_shift}) is trivially fulfilled for any unpaired edge.
The shift condition (\ref{p_shift}) states that the vector elements corresponding to directed edges associated with the same undirected edge must always be equal for common eigenstates.

Analogous steps turn shift conditions for p-attractors (\ref{shift_p_attractors}) into equations for matrix elements of a possible p-attractor $X$
\begin{align*}
X^{k_1(i)}_{k_2(j)} &= X^{l_1(i)}_{l_2(j)}\,, ~~{\rm for~all~}~i,j\in E^{(d)}, K_1,K_2,L_1,L_2 \in 2^E.
\end{align*}
Taking into account all these configurations it simplifies further as
\begin{align}
\label{full_shift_cond}
X^{i}_{j} &= X^{\tilde{i}}_{j} = X^{i}_{\tilde{j}} = X^{\tilde{i}}_{\tilde{j}},
\end{align}
valid for all paired and unpaired edges $i$ and $j$. For general attractors the shift condition (\ref{shift_p_attractors}) must be fulfilled only for configurations $K_1=K_2$ and $L_1=L_2$ and it implies a weaker condition for matrix elements of attractors
\begin{align}
X^{k(i)}_{k(j)} &= X^{l(i)}_{l(j)}\,, ~~{\rm for~all~}~i,j\in E^{(d)}, K,L \in 2^E.
\label{attractor_shift_condition}
\end{align}
While for $j\neq i$ and $j\neq \tilde{i}$ we receive the same set of equations (\ref{full_shift_cond}), for $i=j$ the shift condition reduces to
\begin{align*}
X^{i}_{i} = X^{\tilde{i}}_{\tilde{i}}
\end{align*}
and for $j = \tilde{i}$ to
\begin{align*}
X^{i}_{\tilde{i}} = X^{\tilde{i}}_{i}.
\end{align*}
This is due to the fact that one edge cannot be simultaneously open and closed in one configuration.

Let us conclude this part with summarizing a recipe for finding all attractors. First, we have to determine all common eigenstates (see section \ref{sec:p-Attractors}). Linear combinations (\ref{p_attractors_construction}) provide all p-attractors. The remaining attractors (non-p-attractors) must violate the equality
\begin{align}
\label{broken_shift_cond}
X^{i}_{i} = X^{i}_{\tilde{i}}
\end{align}
at least for one edge $i$. Thus relaxing this condition allows to find all non-p-attractors and complete the set of attractors. The last step is not trivial and differs from case to case, see e.g. our approach for percolated Grover quantum walks on graphs with maximal degree three for two types of shift operators provided in appendix \ref{appendix_non_p}. Note also that the condition (\ref{broken_shift_cond}) can be used to decide whether a given attractor is a p-attractor.



\section{Restricted Percolation}
\label{section4}

Until now, we have assumed that the configuration of open edges can be any subset of $E$. All configurations $K\subset E$ might appear in dynamically percolated quantum walk with a certain probability $\pi_K$. We call this the \textit{full percolation}. On the contrary, scenarios with only a subset of
configurations allowed are called a \textit{restricted
percolation}. We show, that the
asymptotic dynamics of the dynamically percolated quantum walk with a restricted percolation is in many cases the same.

The asymptotic evolutions of dynamically percolated quantum walks with the full percolation and with some restricted percolation are the same if their asymptotic spectra and associated attractors are the same, i.e. if solutions of their attractor equations (\ref{attractors}) coincide. In turn it reduces to the equivalence of their shift conditions (\ref{shift_cond}).
Indeed, even if the empty configuration is not allowed, we can choose any other allowed configuration in the first step of the search for attractors. Or we can use the forbidden empty configuration  $K=\emptyset$ in the calculation, provided the equivalence of their shift conditions was proven first.

The shift condition imposes equality of some matrix elements based on the simultaneous presence/absence of two edges in configurations $K$ and $L$. Clearly, the restricted percolation must contain pairs of configurations capable to restore the same shift condition (\ref{attractor_shift_condition}).
Thus the equivalence of shift conditions requires that for every
pair of edges there is a pair of configurations, where the presence of
both edges changes and there are two pairs of configurations where only the presence of one edge changes. It is easily seen that this holds for a restricted percolation if for every pair of undirected edges it contains three out of the four configurations: both edges present, both edges missing, only the first edge present, only the second edge present.

Let us give some examples of restricted percolation schemes with the same asymptotic behavior as the dynamically percolated quantum walk with the full percolation:
\begin{itemize}
    \item A weakly connected system, in which only configurations with just one open edge are possible. Here, the number of configurations is reduced from $2^{\#E}$ to just $\#E$. (For the asymptotic equivalence we assume that the structure graph has at least 3 edges.)
    \item A system with small perturbations, where only configurations with just one edge closed are allowed.
    \item A system with closed vertices, where a closed vertex means that all its adjacent edges are closed. This kind of restricted percolation will also be asymptotically equivalent to the full percolation when assuming at least 3 vertices.
\end{itemize}
An important property is that adding arbitrary configurations to
schemes which are asymptotically equivalent with the full percolation (e.g. the examples listed above), does not alter their resulting asymptotic evolution. The longtime dynamics of such quantum walks is identical. Note, however, that even though the resulting asymptotic behavior is the same, the rate of convergence towards the asymptotic regime may significantly differ and depends on the actual details of available configurations and their probabilities.


\section{Grover Quantum Walks on simple planar graphs with maximal degree $3$}
\label{sec:3-regular graphs}
The asymptotic behavior of dynamically percolated quantum walks has already been investigated for walks on line graphs \citep{asymptotic1} and square
lattices \citep{asymptotic2}. We now use our general formalism to investigate dynamically percolated quantum walks on graphs with more complex geometries. In particular, we investigate quantum walks on finite simple planar structure graphs with the maximal degree $3$. This class of graphs contains arbitrary cuts of the honeycomb lattice (occurring naturally as graphene) and all its spatial derivatives consisting of hexagonal and pentagonal faces (fullerenes, various carbon nano-tube structures) and many other graphs of interest as for example the graph of the cube, which
will be investigated in detail later.

To keep the same coin at all vertices, we increase the degree in every vertex of the state graph to $3$ by adding unpaired loops. As the structure graph is simple, the added unpaired loops are the only loops in the state graph.

\subsection{The Grover coin}
Since the state graph is designed to be $3$-regular, all the vertex subspaces are $3$-dimensional. As the coin, we use the Grover matrix defined using $\vert \phi\rangle =
(\vert 0\rangle + \vert 1\rangle + \vert 2\rangle )/\sqrt{3}$ as
\begin{align*}
C_v \equiv G_3 &= 2 \vert \phi\rangle\langle \phi \vert - I
=\frac{1}{3}\left[
\begin{array}{rrr}
 -1 & 2 & 2 \\
 2 & -1 & 2 \\
 2 & 2 & -1 \\
\end{array}
\right]
\end{align*}
at all vertices $v\in V$.

The Grover matrix is very convenient for its symmetry. Since it is a linear combination of the identity matrix and a matrix with all elements the same, both of which commute with all permutation matrices, the Grover matrix itself commutes with all permutation matrices. Therefore, the equality $Q G_3 Q^\dagger = G_3$ holds for any permutation $Q$ implying that reordering of base states has no effect. This allows us to investigate large classes of structure graphs simultaneously.

\subsection{Shift operators}
\begin{figure}
    \centering
    \includegraphics[width=150 pt]{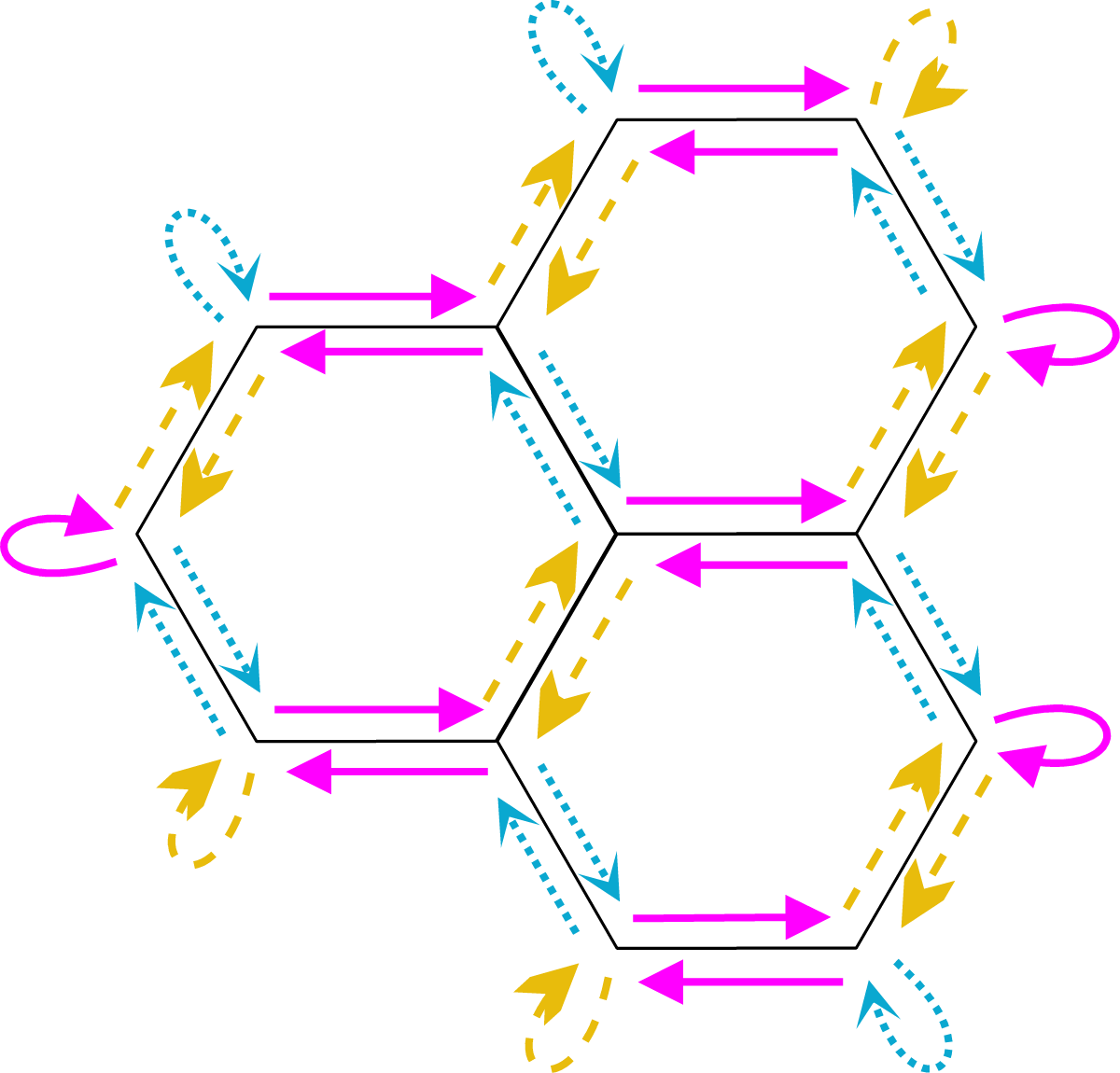}
    \caption{(color online) An example of the action of the reflecting shift operation $R$ on a finite part of a honeycomb lattice. Colors and line types indicate the action of $R$.}
    \label{fig:reflecting}
\end{figure}

With more complex geometries different choices of shift operators arise naturally. Let us discuss some examples. The first obvious case available on any graph is the \textit{reflecting walk}, where $P=I$ and $S=IR=R$. The
action of the reflecting shift operator on a small honeycomb graph is shown in Fig. \ref{fig:reflecting} - the walker jumps back and forth on one undirected edge.

Another interesting shift operators arise if we use \textit{cyclic local permutations} - either clockwise $P^{CW}_v$ or counter-clockwise $P^{CCW}_v$. We consider a general case, in which $P^{CW}_v$ and $P^{CCW}_v$ can be chosen arbitrarily for each vertex $v\in V$. Note that the actions of $P^{CW}_v$ and $P^{CCW}_v$ are already determined once we place our graph into a plane.

\begin{figure}
    \centering
    \includegraphics[width=150 pt]{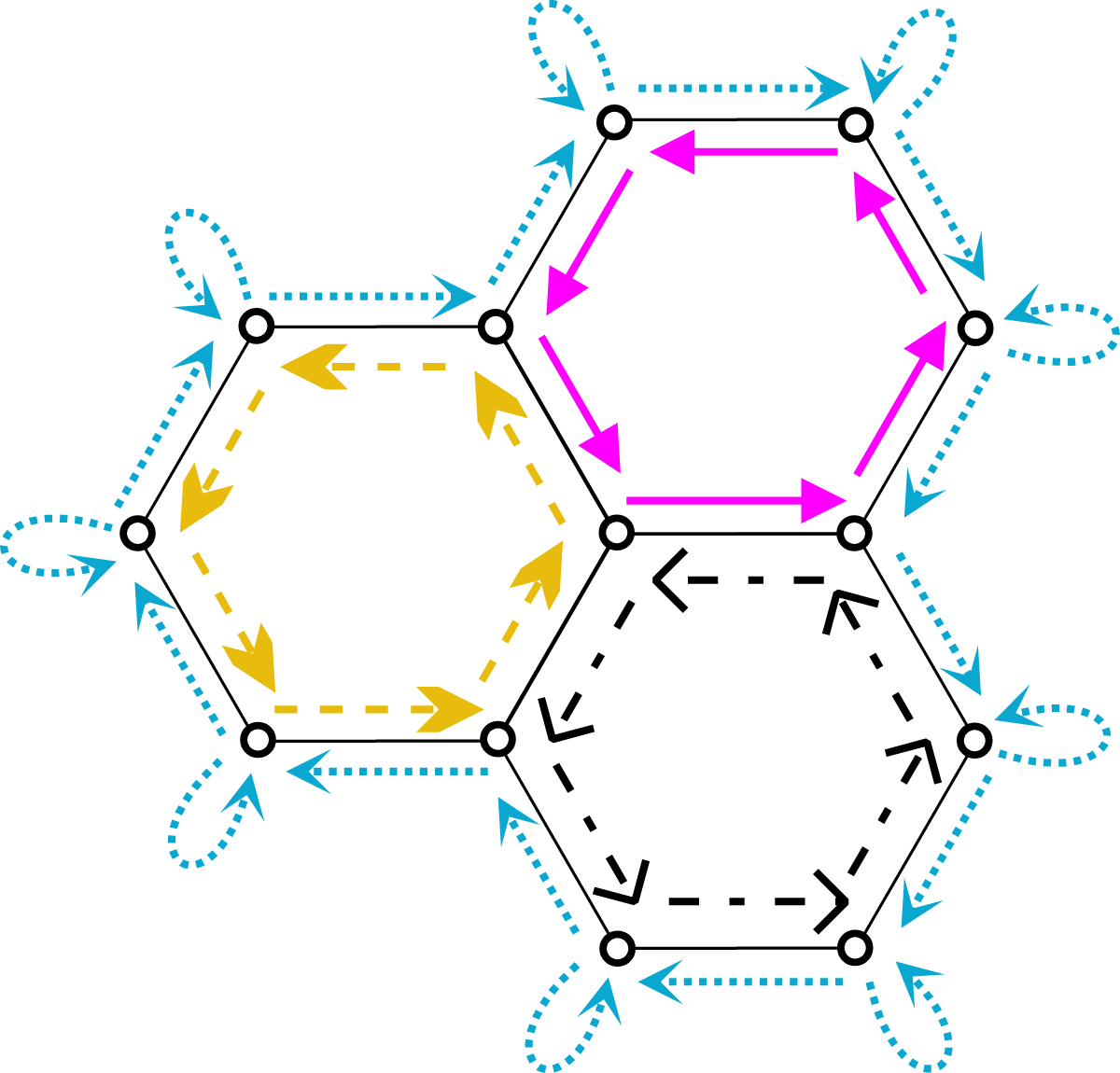}
    \caption{(color online) The action of the cyclic shift operator (the local permutation is $P_v^{CW}$ at all vertices) on a finite part of a honeycomb lattice. The action of $S$ is depicted by colors and line types. On the border, the walker traversing an edge is mapped to the loop and from the loop to the non-loop edge.}
    \label{fig:cyclic}
\end{figure}

\begin{figure}
    \centering
    \includegraphics[width=180 pt]{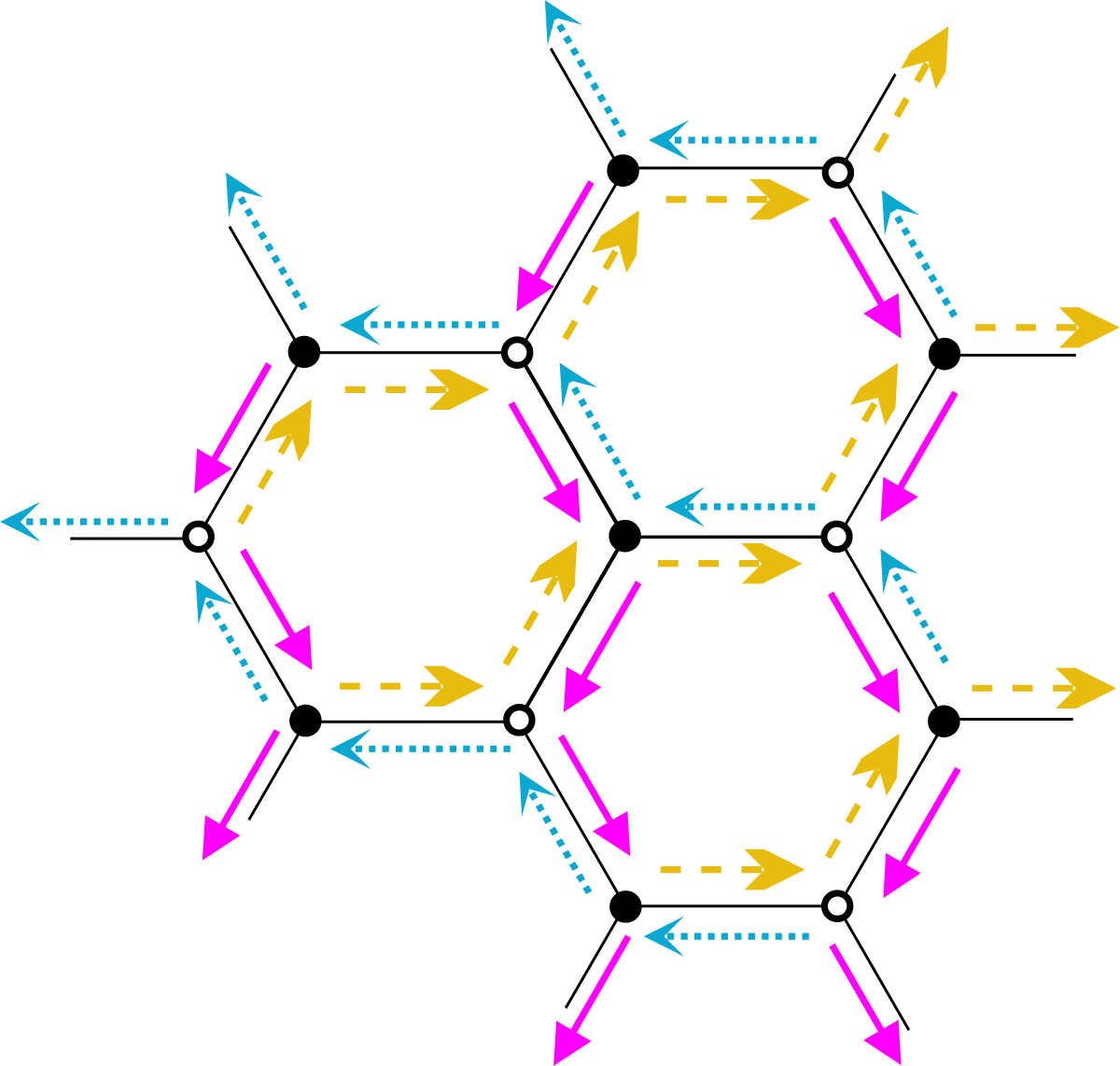}
    \caption{(color online) The action of the transporting shift operator on a honeycomb graph. For better illustration, the action of the operator is demonstrated on a part of a larger graph (border of the graph is not shown). The local permutation is the clockwise rotation $P_v^{CW}$ in one half of the vertices (black) and the counter-clockwise rotation $P_v^{CCW}$ in the other half (white).}
    \label{fig:transporting}
\end{figure}

Honeycomb lattices constitute illuminating examples, where the shift operator maintaining walker's direction is not available. However, we can use shift operators analogous to those shown already in Fig. \ref{fig:different_shifts_square}. The hexagonal variant for the \textit{cyclic} walk is shown in Fig. \ref{fig:cyclic} and for the \textit{transporting} walk in Fig. \ref{fig:transporting}.

While the cyclic walk can be defined on any planar graph, the transporting shift operator requires the graph to be bipartite (see appendix \ref{app:graph theory}). On different graphs, different distributions of $P^{CW}_v$ and $P^{CCW}_v$ may be of a particular interest, e.g. the one in Fig. \ref{fig:dodecahedron}. Note that in all cases the formalism defines the behaviour on the border of a finite graph consistently.

\subsection{Asymptotic behavior of percolated Grover quantum walks with two types of shift operators}
\label{section_asymptotic}
Our main aim is to find asymptotic dynamics of a dynamically percolated quantum walk. This is essentially equivalent finding of attractors satisfying equations (\ref{as_state}). In this section we present a general construction of attractors for the percolated Grover walk on an arbitrary simple planar graph with the maximal degree $3$ for two important types of shift operators: the reflecting shift operator and shift operators with any combination of cyclic local permutations. According to the recipe given at the end of section \ref{sec_The Shift_Condition} we proceed in two steps. Using common eigenstates we first find all p-attractors and as the second step missing non-p-attractors are found. However, both these steps are rather technical and must be performed separately for different shift operators. Our intention is to keep clarity of the presented attractor constructions in the main text and therefore we leave all unnecessary technical details of proofs for appendices \ref{app:asymptotics-proofs}, \ref{appendix_basis}, and \ref{appendix_non_p}.

\subsubsection{Reflecting shift operator}
\label{asymptotic_reflecting}
Our first task is to construct all common eigenstates. Those are all walker's states $\ket{\phi}=\sum_{j\in E^{(d)}}\phi_j\ket{j}$ which simultaneously satisfy the shift condition (\ref{p_shift}) and the eigenvalue\GREEN{coin} condition (\ref{p_one}) for $P=I$ and the Grover coin. The shift condition implies that vector elements (or parameters) $\phi_j$ associated with the same undirected edge in the structure graph must be the same.

Now let us explore the additional constraints imposed by the condition (\ref{p_one}). In each $3$-dimensional subspace $\sH_v$ this condition reads
\begin{align}
\label{g3_eigenvalues}
G_3 \ket{\phi_v} &= \alpha \ket{\phi_v}.
\end{align}
The spectrum of the Grover matrix contains only $1$ and $-1$. The eigenvector associated with the eigenvalue $1$ is
\begin{align}
\label{phi1}
\ket{\phi_v^1} =
\left[
\begin{array}{c}
 1 \\
 1 \\
 1 \\
\end{array}
\right],
\end{align}
which together with the shift condition immediately implies that there is just one common eigenstate associated with the eigenvalue $1$. It is the state $\ket{\phi}$ whose all vector elements are the same.

The analysis of common eigenstates associated with the eigenvalue $-1$ is more involved. Eigenvectors of the Grover matrix corresponding to $-1$ form a two-dimensional subspace orthogonal to $\ket{\phi_v^1}$. Thus, we obtain a particular form of the coin condition, which states that the sum of all parameters $\phi_j$ corresponding to outgoing edges of any vertex must be zero.

Let us first perform a dimensional analysis of the common eigenstates subspace. The dimension of the whole walker's Hilbert space is $3\# V$. The restriction to the subspace of common eigenstates associated with the eigenvalue $-1$ is given by $\# V$ coin conditions and $\# E$ shift conditions. If all these linear constraints are independent, we have $2\#V-\#E$ common eigenvectors associated with the eigenvalue $-1$. It can be shown that these equations are linearly dependent if and only if the structure graph is bipartite and 3-regular (there are no unpaired loops in the state graph). However, after removing any one of the equations, the remaining set is linearly independent. For the proof of both of these statements see appendix \ref{appendix_independence}. Hence we can conclude that the number of linearly independent common eigenstates for the eigenvalue $-1$ is either $N = 2\#V-\#E+1$ for bipartite $3$-regular structure graphs or $N = 2\#V-\#E$ in all other studied cases.

In the following we show an explicit construction of common eigenstates. Proofs of independence and completeness of the chosen basis as well as a detailed description of the common eigenstates and a derivation of their form are given in appendix \ref{appendix_basis}. Here we only provide a selfcontained short guideline how to construct them.

Compared to the situation with the eigenvalue $1$, in this case common eigenstates for the eigenvalue -1 can have zero vector elements. We exploit this fact to construct a basis of common eigenstates from states having limited support. We say that the support of walker's state is limited to some part of the undirected structure graph (e.g. a path), if only vector elements $\phi_j$ corresponding to paired directed edges associated with the undirected edges in question are nonzero.

\begin{figure}
	\centering
	\includegraphics[width=160 pt]{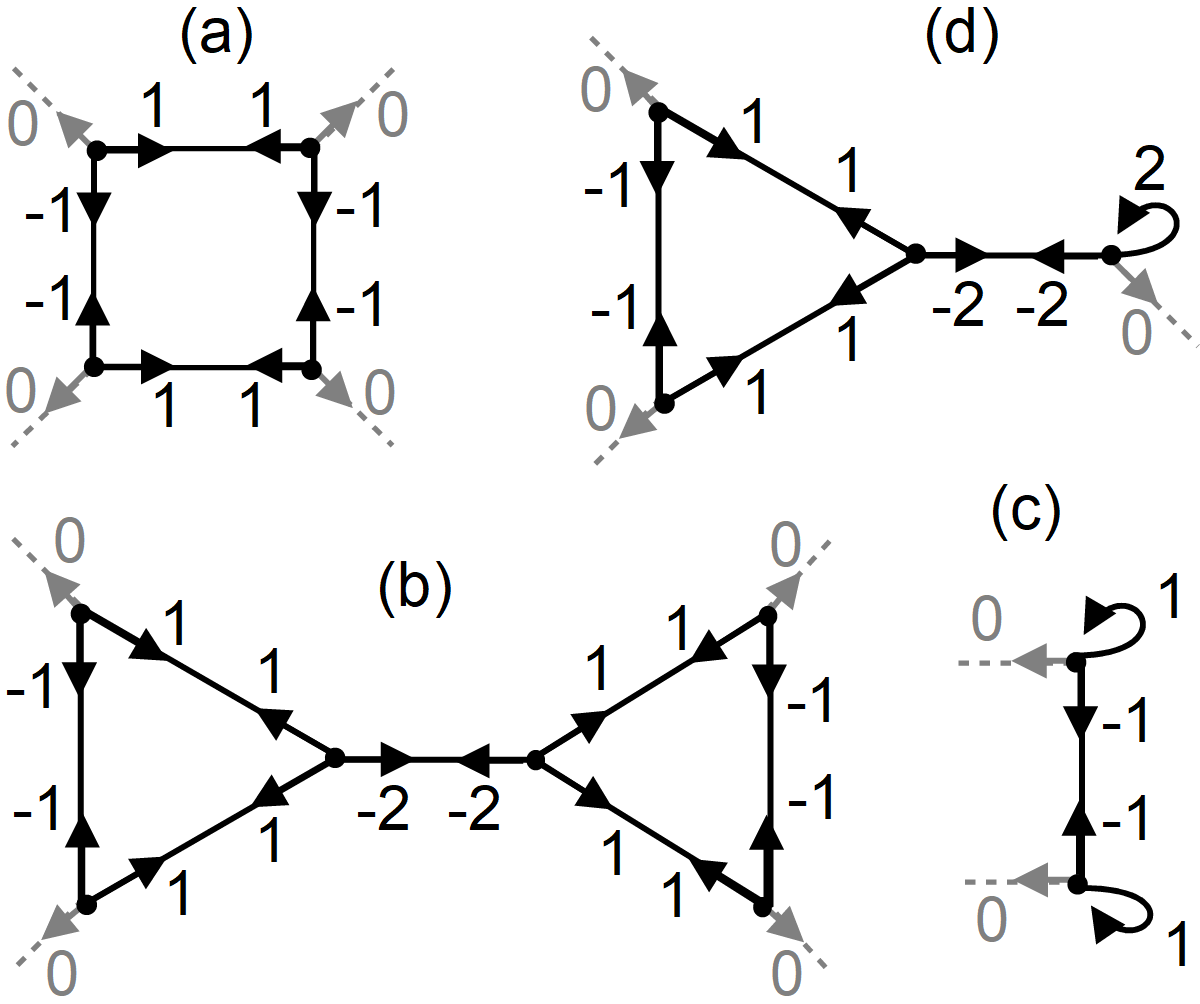}
	\caption{The four basic types of possible common eigenstates for the eigenvalue -1: (a) A-type, (b) B-type, (c) C-type and (d) D-type. Dashed lines represent an arbitrary continuation of the graph where all the corresponding vector elements of the given state are 0.}
	\label{fig:trapped_construction}
\end{figure}

A planar graph drawn into a plane separates it into faces which are either odd-edged or even-edged. Based on that we introduce four types of common eigenstates. Prominent examples of all four types are presented in Fig. \ref{fig:trapped_construction}.  The figure also shows how vector elements of common eigenstates are set along edges for each type of a common eigenstate. The A-type common eigenstate has support limited to one even-edged face. The support of a B-type common eigenstate connects two odd-edged faces by a path. A common eigenstate of the C-type is supported on a path with two unpaired loops on ends. Finally, the D-type are common eigenstates with their supports limited to one odd-edged face connected by a path to one loop. Obviously all these states follow the coin condition and the shift condition.

Now, how to choose these common eigenstates in order to obtain their basis depends on the actual state graph. Note that states differing only in the connecting path would be linearly dependent in the resulting basis of common eigenstates. Thus we always use only one (arbitrarily chosen) state for every face-face, loop-loop or face-loop pair. We meet one of the three situations:
\begin{itemize}
\item[(1)] A state graph without unpaired loops: We use all the A-type common eigenstates and add B-type common eigenstates connecting one arbitrarily chosen and fixed odd-edged face with all the other odd-edged faces.
\item[(2)] A state graph with unpaired loops and a structure graph with only even-edged faces: We take all the A-type common eigenstates and complete the basis by C-type common eigenstates connecting one arbitrarily chosen and fixed loop to all the others.
\item[(3)] A state graph with unpaired loops and a structure graph with some odd-edged faces: We use the same common eigenstates as for the first situation and add one D-type common eigenstate for every loop (with the odd-edged faces chosen at will).
\end{itemize}
In all three cases we avoid employing the outer face (the rest of the plane outside the graph) for the construction of states as it would result in a linearly dependent set of common eigenstates.
It is likely that due to some symmetry of the graph, different choices for the basis of common eigenstates might be more convenient. Nevertheless, the procedure above is guaranteed to provide a complete and linearly independent set, from which all other common eigenstates can be obtained.

Altogether, for an arbitrary planar graph with maximal degree 3 we have found one common eigenstate corresponding to the eigenvalue $1$ and $N$ common eigenstates corresponding to $-1$. That results in $2\times N$ p-attractors corresponding to the eigenvalue $-1$ and $N^2+1$ p-attractors corresponding to $1$.

The last step is to find the remaining attractors - the
non-p-attractors. This step is obviously as important as the search
for p-attractors to obtain the complete asymptotic dynamics (\ref{as_state}). Using a novel method we prove that for the percolated reflecting Grover quantum walk on an arbitrary simple graph with maximal degree $3$ we only need to add the identity operator, which corresponds to the eigenvalue 1. This proof does not use the planarity assumption so it applies to all structure graphs with maximal degree $3$. The proof itself is moved to appendix \ref{appendix_non_p_reflecting} because it is technically difficult and has not much further relevance to the remaining parts of the text.

\subsubsection{Shift operators with cyclic local permutations}
In this part we consider a walk with the Grover coin and cyclic local permutations $P^{CW}_v$ and $P^{CCW}_v$ distributed among vertices. Similarly to the case with the reflecting shift operator, it can be shown that apart of p-attractors the only needed non-p-attractor is the identity operator. A detailed proof of this claim is left for appendix \ref{appendix_non_p_cyclic}. Thus, in the following we focus our analysis on common eigenstates and corresponding p-attractors. The one-vertex restriction of the equation (\ref{p_one}) has two variants
\begin{align}
\label{eq_coin_local_permutations}
G_3 P^{CW}_v\ket{\phi_v} &= \alpha \ket{\phi_v}, \\
G_3 P^{CCW}_v\ket{\phi_v} &= \alpha \ket{\phi_v}. \nonumber
\end{align}
Both eigenvalue conditions (\ref{eq_coin_local_permutations}) have the same spectrum. The eigenvalues are $\alpha_1=1, \alpha_2=\e^{i\pi/3}$ and $\alpha_3=\e^{-i\pi/3}$. For $\alpha_1=1$ we have an eigenvector of the form (\ref{phi1}). Up to an overall phase, vector elements of each of the remaining eigenvectors can have only three possible values. We denote them by red, green and blue color: $r=1$, $g=\e^{i\frac{2\pi}{3}}$, $b=\e^{-i\frac{2\pi}{3}}$. These are in a clockwise order in the eigenvector for $P^{CW}_v$ corresponding to $\alpha_2=\e^{i\pi/3}$ or for $P^{CCW}_v$ corresponding to $\alpha_3=\e^{-i\pi/3}$ and in a counter-clockwise order otherwise. The order of colors prescribed by the local permutation to each edge of the state graph gives us a convenient form of the coin condition (\ref{eq_coin_local_permutations}) at each vertex.

The shift condition (\ref{p_shift}) requires that the vector elements corresponding to one undirected edge are the same in a common eigenstate.  Again, we find that there is always one common eigenstate corresponding to the eigenvalue $1$ having all elements equal. The existence of common eigenstates for the eigenvalues $\alpha_2=\e^{i\pi/3}$ and $\alpha_3=\e^{-i\pi/3}$ is not guaranteed in general, but it can be checked by a straightforward coloring procedure. We first assign color (value $r$, $g$ or $b$) to one arbitrary edge in the state graph. Obviously, this only fixes the irrelevant overall phase of a possible non-stationary common eigenstate. Then the coin condition determines colors of edges outgoing from this vertex and together with the shift condition it induces the coloring in other vertices. This process either results in a consistently edge-3-colored state graph representing a common eigenstate or we encounter one of the two types of conflict. Two edges originating in one vertex can have the same color or there can be different colors for two paired edges corresponding to the same undirected edge. In these cases the common eigenstate for that eigenvalue does not exist for the given distribution of cyclic local permutations. Actually, if a non-trivial common eigenstate exists for one eigenvalue, the existence of the other one for the conjugate eigenvalue is guaranteed as well. Indeed, swapping directions of all local permutations is equivalent to "looking" at the plane with the graph from the other side, which can not create or remove a conflict of shift and coin conditions. We formulate these findings in the following theorem.

\begin{theorem}
\label{theorem1}
For a given structure graph (with maximal degree 3), a corresponding 3-regular state graph and a distribution of cyclic local permutations at vertices there exist common eigenstates for eigenvalues $\alpha_2=\e^{i\pi/3}$ and $\alpha_3=\e^{-i\pi/3}$ in the dynamically percolated Grover quantum walk if and only if the coloring procedure described above results in a consistent edge-3-coloring of the state graph.
\end{theorem}

The common eigenstates, if they exist, are unique for a given eigenvalue and their vector elements $\phi_j$ are given by colors of the corresponding edges. Thus, since all the three colors correspond to non-zero values, there is no common eigenstate with a support limited to some part of the state graph.

\begin{figure}
    \centering
    \includegraphics[width=220 pt]{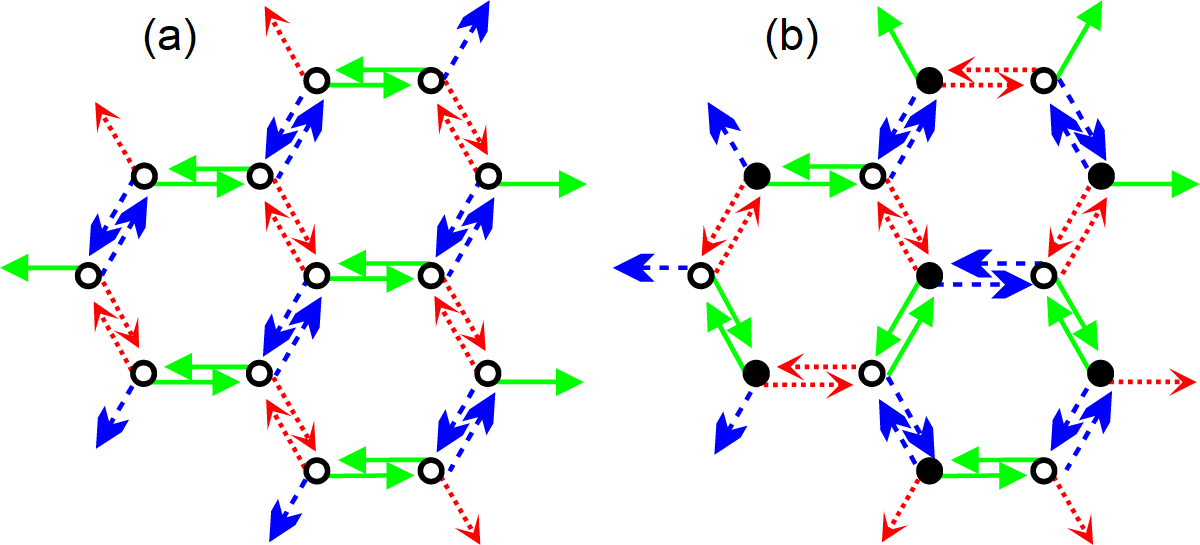}
    \caption{Construction of non-stationary common eigenstates for percolated quantum walks on the honeycomb lattice: (a) the common eigenstate for the cyclic walk corresponding to eigenvalue $\alpha_2$ and (b) the common eigenstate for the transporting walk corresponding to eigenvalue $\alpha_2$. The non-filled vertices represent local permutations $P^{CW}_v$ and filled vertices local permutations $P^{CCW}_v$. The border edges can either represent undirected loops or a continuation of the graph.}
    \label{fig:coloring_honeycomb}
\end{figure}

Let us apply this result to some examples. From the coloring procedure we see that both cyclic and transporting percolated quantum walks introduced for the honeycomb lattice have the non-trivial common eigenstates. The common eigenstates for the cyclic walk are represented by a coloring where all parallel\footnote{Here we mean "parallel" in the geometry of the honeycomb lattice, not in the graph theory terminology.} edges have the same color (see Fig. \ref{fig:coloring_honeycomb} (a)) and for the transporting walk the coloring has only two alternating colors in every hexagon (see Fig. \ref{fig:coloring_honeycomb} (b)).

It is of relevance to consider also the reverse problem. Assume that a structure graph and its corresponding  state graph are given. Is there a distribution of cyclic local permutations resulting in the presence of common eigenstates corresponding to eigenvalues different from 1? The next corollary of the theorem \ref{theorem1} provides a complete answer.

\begin{corollary}
\label{theorem2}
For a given structure graph (with maximal degree 3) and its corresponding 3-regular state graph there exists a distribution of cyclic local permutations resulting in the presence of common eigenstates associated with eigenvalues $\alpha_2=\e^{i\pi/3}$ and $\alpha_3=\e^{-i\pi/3}$ for the dynamically percolated Grower walk if and only if there exists an edge-3-coloring of the undirected structure graph.
\end{corollary}

The first part of the proof is obvious. According to theorem \ref{theorem1}, the existence of a non-stationary common eigenatate implies an edge-3-coloring of the state graph and this implies an edge-3-coloring of the structure graph. Let us consider the reverse implication. We color the paired edges by the colors of the corresponding undirected edges, which ensures fulfilling the shift condition. Colors of unpaired loops are also given for vertices with only one unpaired loop. There is an ambiguity only at vertices with two unpaired loops. We pick one possible option. Now, the coloring of the state graph represents a common eigenstate and after we choose whether it corresponds to $\alpha_2$ or $\alpha_3$, the coin condition determines the required distribution of cyclic local permutations.

\begin{figure}
    \centering
    \includegraphics[width=150 pt]{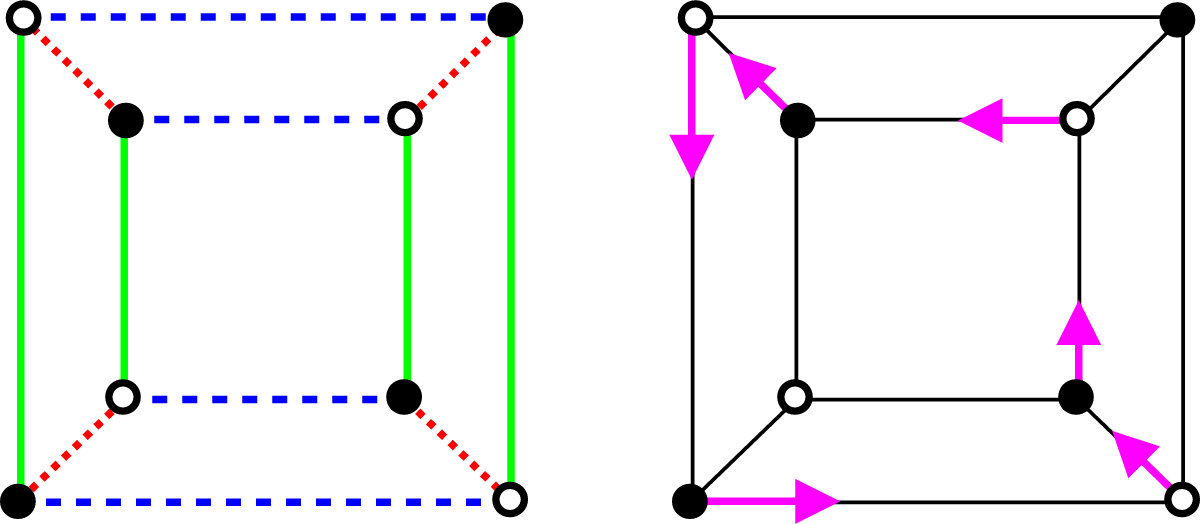}
    \caption{(color online) (a) The edge-3-coloring representing the common eigenstate corresponding to eigenvalue $\alpha_2$ and (b) the action of the corresponding "transporting" shift operator on the cube graph. The shift operator generates four cycles in the state graph; only one of them is depicted here. The non-filled vertices represent local permutation $P^{CW}_v$  and filled vertices local permutation $P^{CCW}_v$.}
    \label{fig:cube_both}
\end{figure}

Theorem \ref{theorem2} applies to some well known graph structures. In Fig. \ref{fig:cube_both} we can see that there exists an edge-3-coloring of the cube graph corresponding to a shift operator analogous to the transporting walk on the honeycomb lattice. We decide to use this coloring as an eigenstate corresponding to $\alpha_2$, which results in the distribution of local permutations depicted in the figure. On contrary, the cyclic walk on the cube graph has only the common eigenstate corresponding to the eigenvalue $1$.

\begin{figure}
    \centering
    \includegraphics[width=220 pt]{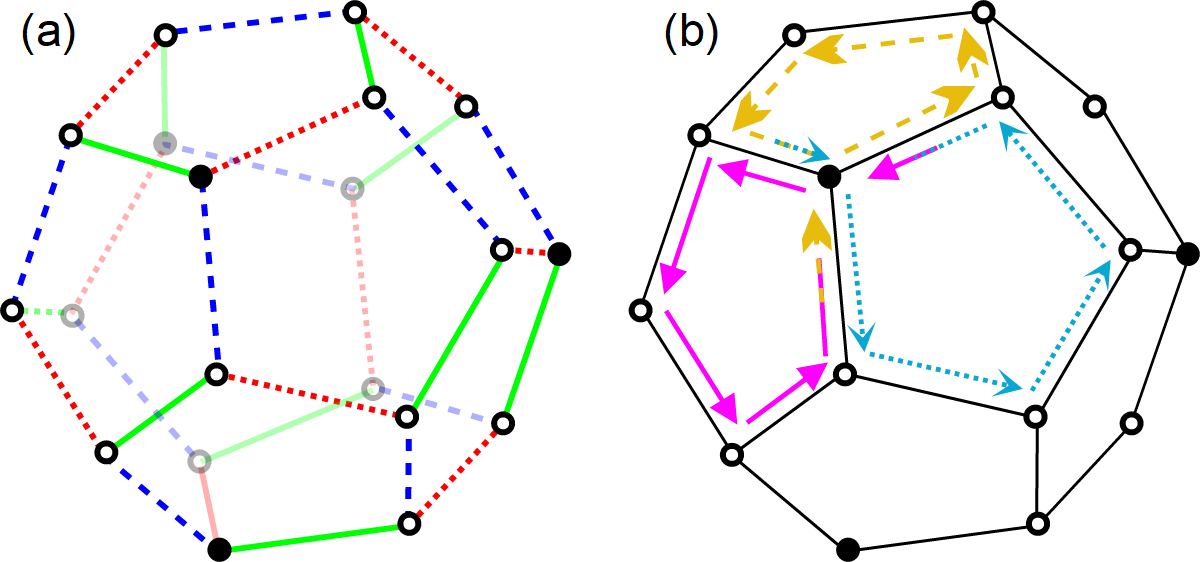}
    \caption{(color online) (a) The edge-$3$-coloring of the dodecahedron resulting in a distribution of cyclic local permutations $P^{CW}_v$ (non-filled points) and $P^{CCW}_v$ (filled points). (b) The action of the corresponding cyclic shift operator. The shift operator generates four cycles in the state graph, only one cycle is depicted here by colors and line types of the arrows.}
    \label{fig:dodecahedron}
\end{figure}

In order to provide an example of a more complex graph, the Fig. \ref{fig:dodecahedron} presents an edge-3-coloring of the dodecahedron and the action of the corresponding shift operator. Hence, due to the theorem \ref{theorem2}, this dynamically percolated Grower walk has non-stationary common eigenstates.

\begin{figure}
    \centering
    \includegraphics[width=160 pt]{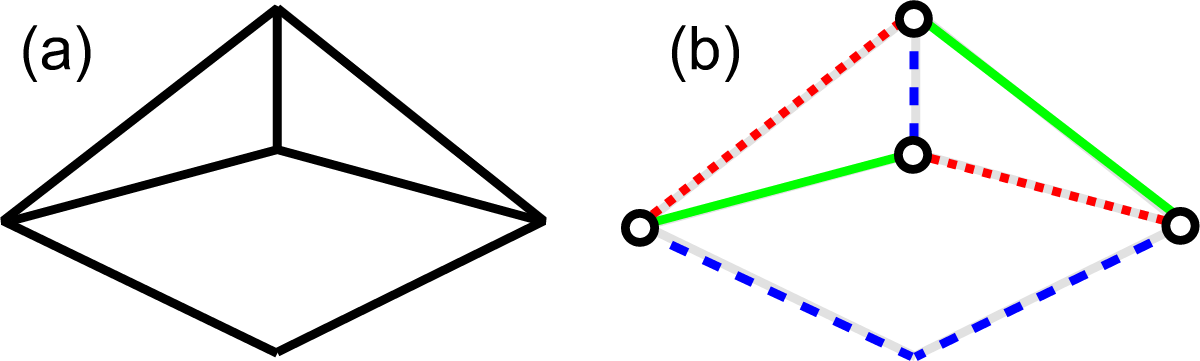}
    \caption{(color online) (a) A planar graph with maximal degree $3$, which is not edge-3-colorable. (b) An example of a wrong edge coloring.}
    \label{fig:non_colorable}
\end{figure}

Naturally, there are graphs for which the reverse task has no solution. Already some planar graphs with maximal degree 3 do not allow any edge-$3$-coloring. An example of such graph is shown in Fig. \ref{fig:non_colorable}, where the nonexistence of edge-3-coloring can be checked by trying every vertex as a starting point of the coloring procedure.


\section{Excitation transfer in reflecting Grover walk}
\label{section6}

Quantum walks constitute a well designed and frequently employed tool suitable to model and investigate various transport scenarios through different media. One of the possible applications is the study of excitation transfer from a source to a sink and in particular its efficiency. The role of the excitation and the medium is taken by the walker and the underlying graph of the quantum walk respectively. The walker is initiated in a state whose support typically overlaps only with a few vertices of the graph (the source). Similarly, the sink is a chosen set of vertices, through which the walker leaves the graph. To be explicit, we introduce a \textit{sink subspace} as the sum  of vertex subspaces of vertices from the sink. Let $T$ be the orthogonal projection on the sink subspace, which after each step of the walker's evolution destroys the part of the walker's wave function overlapping with the sink subspace. The rest of the walker's wave function keeps evolving.

If $U$ describes one step of walker's unitary evolution, in the presence of the sink one step of the walker's state dynamics is modified as
\begin{align*}
\ket{\psi(t+1)} = (I-T)U\ket{\psi(t)}.
\end{align*}
Analogously, the dynamics of the percolated quantum walk with the sink is redefined.\footnote{Note that since the sink always covers whole vertex subspaces, the operator $I-T$ commutes with the coin operator and therefore the relation between attractors of both variants $U^{(1)}$ and $U^{(3)}$ is not affected by the introduction of the sink.} The modified dynamics with the sink is not trace-preserving. The actual value of the trace of the walker's density operator, i.e. $p(t)=\Tr(\rho(t))$, expresses the probability that the excitation is still present in the medium. On the other hand, $q(t)=1-p(t)$ gives us probability that the excitation was already transported to the sink. A question of our central interest is the overall efficiency of the whole process, which is the total probability
\begin{equation}
\label{def_efficiency}
q = \lim_{t \rightarrow +\infty} q(t)= 1 - \Tr\left(\lim_{t \rightarrow +\infty} \rho(t)  \right)
\end{equation}
with which the excitation is transferred to the sink. This efficiency (\ref{def_efficiency}) is determined solely by the asymptotic dynamics of the original evolution modified by the presence of the sink. In \citep{assisted_transport} it was shown that the effect of the sink can be easily taken into account. The resulting asymptotic evolution is again expressed in terms of original attractors (\ref{as_state}), but now we employ only those having zero overlap with the sink subspace. Consequently, if supports of all attractors are spread over the whole structure graph, the efficiency of the excitation transfer is one. As we point out in the following part, this is interestingly not always the case.

\subsection{Transfer to a sink and trapped states}
In section \ref{asymptotic_reflecting} we have found that the reflecting Grover walk with dynamical percolation has a rich structure of localized common eigenstates. As illustrated in Fig. \ref{fig:trapped_construction} these states are typically confined in a subset of vertices of the graph. The walker being in such a state stays trapped in some part of the graph for all times and this trapping is not disrupted even by dynamical percolation. Nevertheless, we stress that other localized eigenstates may be present in the dynamics of the non-percolated quantum walk, which are excluded by the percolation, as it is shown for the quantum walk on the cube graph below.

In view of equation (\ref{def_efficiency}), we see that the existence of localized attractors is the only factor responsible for a lower efficiency of the excitation transport. Let us discuss the structure of these attractors in detail. Since the identity operator always overlaps with the sink subspace, only the p-attractors constructed from the localized common eigenstates are possible candidates allowing the walker to stay away from the sink permanently. In the following text we call these localized common eigenstates \textit{trapped states} and their important subset are so-called \textit{sr-trapped states} ("sink-resistant") -- the trapped states orthogonal to the sink subspace.

To evaluate the efficiency (\ref{def_efficiency}) in a percolated quantum walk we need an orthonormal basis of sr-trapped states. As discussed in section \ref{asymptotic_reflecting}, we already
have the needed complete set of trapped states available and we have to filter out the trapped states having a non-zero overlap with the sink. It can be done using the following recipe. We take the base states (corresponding to particular directed edges) of the sink subspace into consideration one by one and follow recipe for one of three possible situations. First, if the base state has zero overlap with all the trapped states in the current set, we just keep the whole set. Second, if it has an overlap with exactly one trapped state, we simply remove this trapped state from the current set. Third, if there are more trapped states overlapping with this sink state, we have to search for all possible linear combinations of trapped states having zero overlap with this sink state. However, also in this case, the dimension of the subspace of sr-trapped states decreases exactly by one due to one added restriction. At the end we are left with the complete set of sr-trapped states. Note that the third possibility can often be avoided by an appropriate choice of the original set of trapped states and by the order in which the sink base states are considered.

\subsection{Trapped states for different shift operators}

The existence of trapped states is closely associated with the choice of the shift operator. While the reflecting Grover walk exhibits trapping, the cyclic walk does not. Consequently, the efficiency of the excitation transport is always one for percolated Grower walk with any cyclic shift operator. On the other hand, as it is presented in section \ref{time_evolution}, a modification of the permutation $P$ defining the shift operator of a quantum walk can be compensated by a corresponding change of the coin. Thus, the cyclic walk investigated before with local permutation $P_v = P^{CW}$ at all vertices with a modified coin
\begin{align*}
G_3 \left(P^{CW}\right)^{-1} &=
\frac{1}{3}
\left[
\begin{array}{rrr}
 2 & 2 & -1 \\
 -1 & 2 & 2 \\
 2 & -1 & 2 \\
\end{array}
\right].
\end{align*}
shows the same behavior as the reflecting walk with the standard Grover coin $G_3$.

\subsection{Example: percolated reflecting Grover quantum walk on a cube}
\begin{figure}
    \centering
    \includegraphics[width=120 pt]{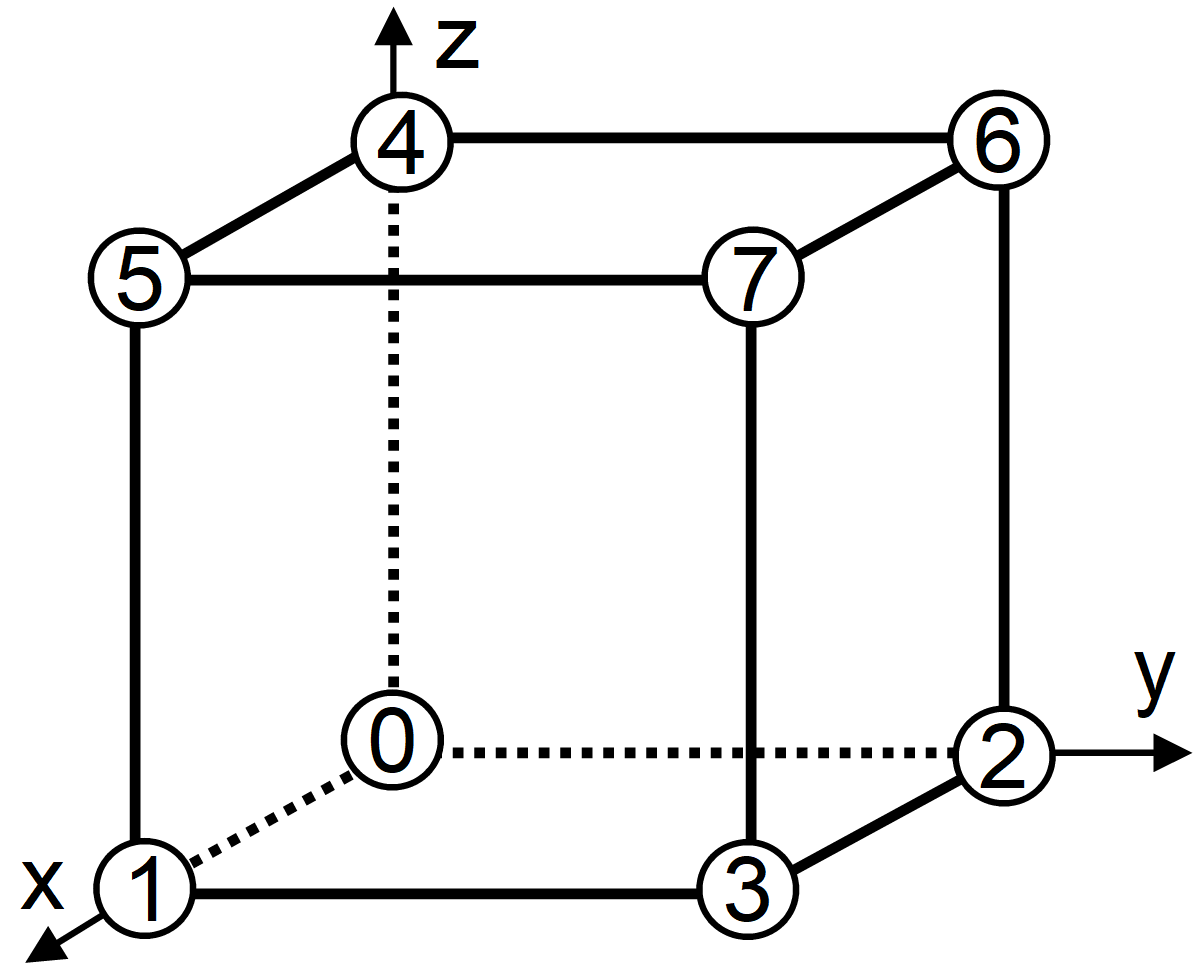}
    \caption{Coordinates on the cube graph. The vertex numbers are chosen so that they correspond to binary numbers given by coordinates $zyx$. We denote faces by their position in this figure so for example the "left"~face has vertices $v_0, v_1, v_5$ and $v_4$.}
    \label{fig:cube}
\end{figure}

One of the simplest examples of $3$-regular graphs is the cube graph. We employ this example to show step by step how to use our theory to construct trapped states and evaluate the efficiency of the excitation transport. Let us position the cube in the coordinate system as shown in Fig. \ref{fig:cube}. Every vertex has one edge in the direction of each axis and we use this to denote walker's states, i.e. the computational basis is chosen in the order $e_x, e_y, e_z$ in every vertex.

The graph is bipartite and has $8$ vertices, $12$ edges and no loops are added in the state graph. Therefore, we must find $N=2\#V-\#E+1=5$ common eigenstates corresponding to the eigenvalue $-1$ (the trapped states). The cube has $6$ even-edged faces. According to the recipe given in section \ref{asymptotic_reflecting}, we simply choose $5$ of those and use common eigenstates of the A-type corresponding to these faces. Just for illustration, the trapped state corresponding to the "left" face reads
\begin{eqnarray}
\label{left_face_state}
|L\rangle &=& |v_0e_x\rangle+|v_1e_x\rangle - |v_1e_z\rangle-|v_5e_z\rangle \nonumber \\
&+& |v_5e_x\rangle+|v_4e_x\rangle - |v_4e_z\rangle-|v_0e_z\rangle.
\end{eqnarray}
For the eigenvalue $1$ there is only one common eigenvector with all its elements equal to one.

The construction described in section \ref{sec:p-Attractors} results in a set of $36$ linearly independent p-attractors, $10$ corresponding to the eigenvalue $-1$ and $26$ corresponding to eigenvalue $1$. To complete the set of attractors we add the identity operator, the only needed non-p-attractor, which corresponds to the eigenvalue 1. This allows us to calculate the asymptotic regime of dynamics for any initial state.

We exploit the obtained attractor structure to study the excitation transfer on the cube. Let the sink be located in the vertex $v_7$ and the initial state of the excitation localized in the opposite vertex $v_0$. To simplify construction of required sr-trapped states, we remove from the $6$ common eigenstates of the type A corresponding to eigenvalue $-1$ the one with a non-zero sink overlap, for example the one corresponding to the "top" face. Following the recipe for sr-traped states, we remove the "right" face state and the "front" face state and we are finally left with $3$ linearly independent sr-trapped states corresponding to the "bottom", "left", and "back" faces. The probability of trapping (complement to the efficiency of the excitation transport) is simply the transition probability from the initial state to the subspace spanned by these three sr-trapped states. To calculate this probability, we need to orthonormalize the set of sr-trapped states numerically.

Depending on the initial state, the efficiency of the excitation transfer ranges from $70~\%$ to $100~\%$. The full transfer occurs for the initial state $\ket{\psi_0}=\frac{1}{\sqrt{3}}(|v_0e_x\rangle+|v_0e_y\rangle+|v_0e_z\rangle)$, which is by construction orthogonal to all trapped states. On the other hand, the efficiency of the transfer can not be arbitrary small, because there is no sr-trapped state localized in the vertex $v_0$. States with the minimum transfer are exactly those orthogonal to $\ket{\psi_0}$. As all the extremal initial states are eigenstates of the Grover coin, the result is the same for both variants $U^{(1)}$ and $U^{(3)}$ generating the evolution. We stress that the analytical investigation of the efficiency is possible due to the knowledge of the analytical form of the trapped states.

For comparison we also investigate (numerically) the efficiency of the excitation transport in the non-percolated version of the reflecting quantum walk on a cube graph. Clearly, the common eigenvectors present in the percolated version are also eigenvectors for the non-percolated walk, so the trapping is again present for most of the initial states. Nevertheless, additional trapped eigenvectors can be identified. There are eigenstates corresponding to the eigenvalue -1, where for every undirected edge one of the values of the corresponding vector elements is $1$ and the other one is $-1$. For the cube we have three more sr-trapped eigenvectors where for example the one corresponding to the "left" face is
\begin{eqnarray}
\label{left_face_state}
|\tilde{L}\rangle &=& |v_0e_x\rangle-|v_1e_x\rangle + |v_1e_z\rangle-|v_5e_z\rangle \nonumber \\
&+& |v_5e_x\rangle-|v_4e_x\rangle + |v_4e_z\rangle-|v_0e_z\rangle.
\end{eqnarray}

\begin{figure}
    \centering
    \includegraphics[width=220 pt]{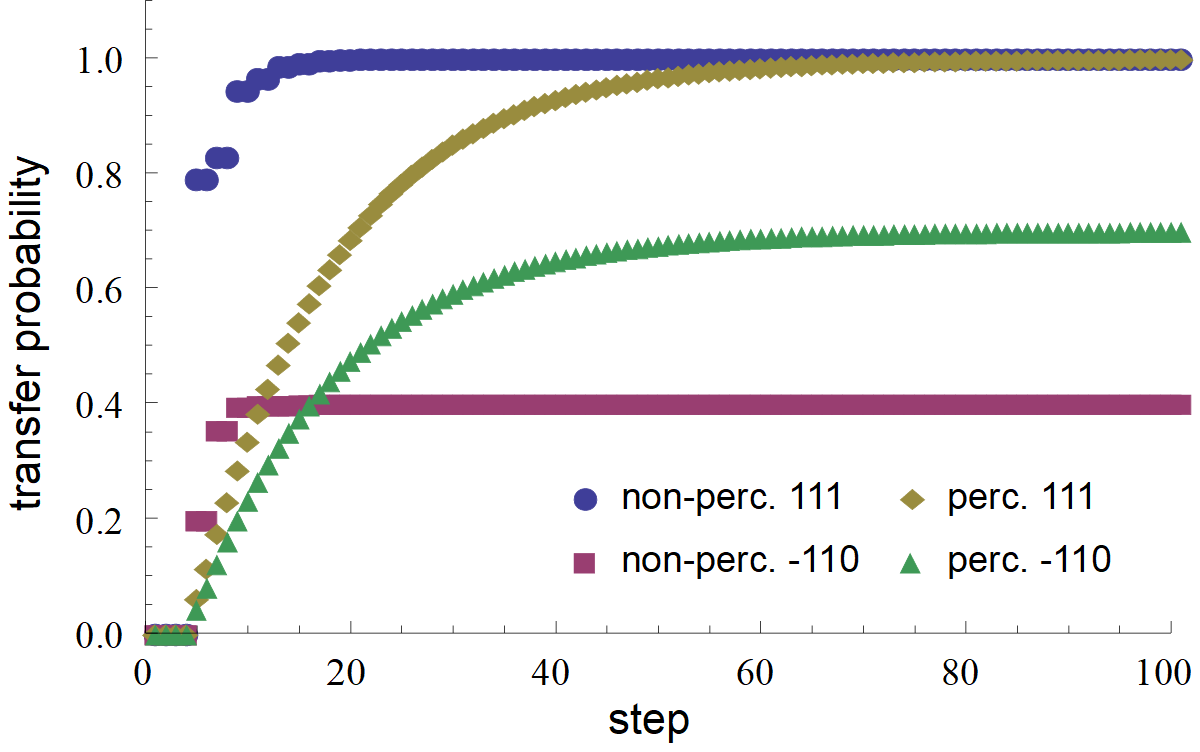}
    \caption{Numerical simulation of the Grover quantum walk on a cube graph with the reflecting shift operator for four situations: without percolation, initial states $\ket{\psi_0}=\frac{1}{\sqrt{3}}(|v_0e_x\rangle+|v_0e_y\rangle+|v_0e_z\rangle)$ (blue circles) and $\ket{\psi_1}=\frac{1}{\sqrt{2}}(|v_0e_x\rangle-|v_0e_y\rangle)$ (purple squares) and with percolation, the same initial states $\ket{\psi_0}$ (yellow diamonds) and $\ket{\psi_1}$ (green triangles). The horizontal axis shows the number of steps and the vertical axis the efficiency of the excitation transfer.}
    \label{fig:cube_all_reflecting}
\end{figure}

The vector $\ket{\psi_0}=\frac{1}{\sqrt{3}}(|v_0e_x\rangle+|v_0e_y\rangle+|v_0e_z\rangle)$ is again orthogonal to all the trapped eigenstates and therefore is fully transferred. For the states orthogonal to $\ket{\psi_0}$, the transfer efficiency is only $40 \%$, so the chance of trapping is doubled compared to the percolated walk. This is given by the presence of the other localized eigenvectors. The results for both dynamically percolated and non-percolated walks are illustrated in Fig. \ref{fig:cube_all_reflecting}.

As the results show, percolation enhances the efficiency of the transfer on the
studied graph by excluding some of the trapped states from the
asymptotic regime. This result obviously applies to other
$3$-regular state graphs, since analogous trapping may be present. Note
also, that the analytical solution of the percolated quantum walk
brings a significant insight into transfer properties of the
non-percolated walk.



\section{Conclusions}
\label{section7}

We studied quantum walks on graphs with nonuniform vertex degree for
different shift operators. To tackle these situations and unify them
to one common framework we introduced an alternative definition
of the quantum walk. The alternative definition respects the main
components exploited for the definition of a discrete time quantum walk. These parts are
typically: the underlying graph (capturing the geometric structure of the
walker's positions and the interconnecting links), the coin (mixing
the internal degree of freedom) and the shift operator (encoding
directions of walker's moves at each vertex in dependence on the the internal state). Altering the definitions of the walker's state space and the shift operator together with the
coin choice of the walk we succeeded to define a broad class of
quantum walks able to cover many physically relevant situations.
It was shown that various quantum walks given by
different shift operators can be mapped onto the one with the
reflecting shift operator and properly adjusted coin operator and
initial state. This allows for a simple classification of different
walker's behavior regarding different shift operators.

Within the developed framework we can easily design percolated
quantum walks on general graphs and analyze their properties in a
convenient way (leading to analytically solvable problems). Such graphs or structures influenced by imperfections naturally occur when we study, for instance,
excitation transfer across large molecules (polymers) or purpose
designed materials with imperfections. For percolated or restricted percolated quantum
walks we identified equations determining the asymptotic space
needed for the construction of the asymptotic dynamics. Using the
pure state ansatz, the construction of the asymptotic space
is worked out in detail for dynamically percolated Grower walks on graphs with maximal degree $3$ (including the
case of a cube, fullerene like structures or finite tree like
structures) driven by two families of shift operators. We did prove that the
found elements of the asymptotic set, after being supplemented by
the identity, form the complete asymptotic space and hence the asymptotic dynamics is fully specified. For percolated Grover quantum walks with the class of cyclic shift operators we showed, that it is possible to choose a cyclic shift operator resulting in non-stationary asymptotic behavior of the walker if and only if there is a edge-3-coloring of the given structure graph.
In addition we found conditions under which the walks exhibit trapping, i.e. states which remain localized and stationary under the action of the walk
evolution operator.  Using the obtained results we gave a simple
application of our approach for the study of overall transfer
efficiency of localized excitation to a prescribed target position
on a cube. The transfer on other graph structures with maximal degree $3$ like
fullerene related planes, tubes or tree like structures are left for a next publication as the relevant results go clearly beyond the scope of the present
paper.

\section*{ACKNOWLEDGEMENS}
J.  N., J. M. and I. J. have been supported by the Czech Science foundation (GA\v CR) project number 16-09824S, RVO14000, by Grant Agency of the Czech Technical University in Prague, grant No. SGS16/241/OHK4/3T/14, and by the project "Centre for Advanced Applied Sciences", Registry No. CZ.02.1.01/0.0/0.0/16\_019/0000778, supported by the Operational Programme Research, Development and Education, co-financed by the European Structural and Investment Funds and the state budget of the Czech Republic.

\appendix

\section{Graph theory basics}
\label{app:graph theory}
In this part we recall several concepts from graph theory used in the main text. We put emphasis on simple and intuitive descriptions since rigorous treatment of the topic can be found in numerous textbooks.

We use curly brackets for unordered sets, so $\{a,b\}=\{b,a\}$, and normal brackets for ordered tuples, so $(a,b)\neq (b,a)$.

An \textit{undirected graph} $G(V,E)$ is a set of vertices $V$ and a set of edges $E$, where every edge connects two vertices and it has no orientation. In contrast, a \textit{directed graph} $G^{(d)}(V,E^{(d)})$ has edges from $E^{(d)}$ with orientation from one vertex towards another. In this work we even use \textit{mixed graphs} $G^{(m)}(V,E,E^{(d)})$ having both undirected and directed edges.

\begin{figure}
    \centering
    \includegraphics[width=160 pt]{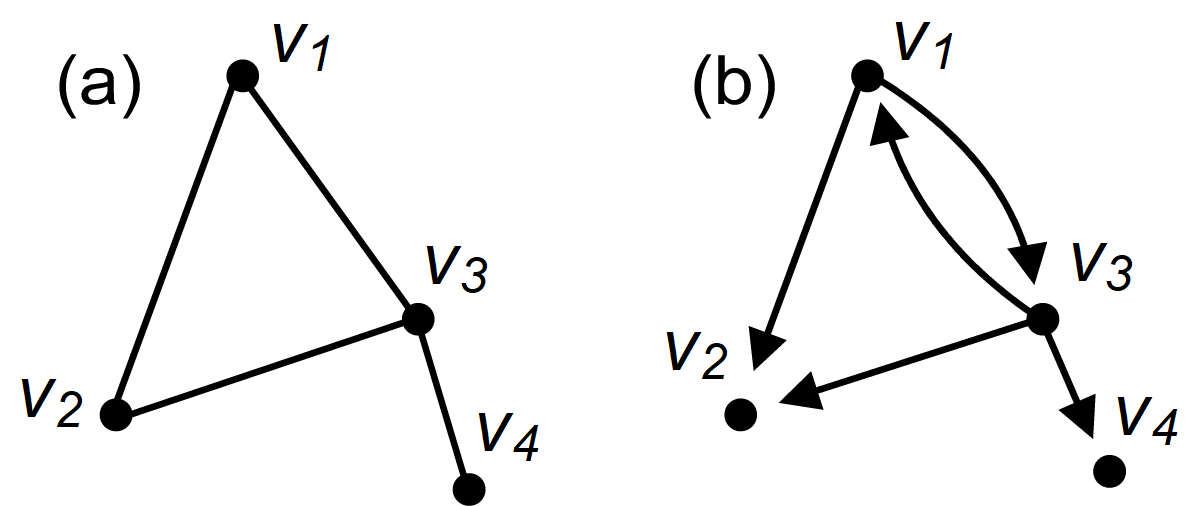}
    \caption{Examples of (a) an undirected graph and (b) a directed graph.}
    \label{fig:graphs_examples}
\end{figure}

While graphs are abstract, they are often represented by an embedding ("drawing") into a plane. A representation of a graph with the set of vertices $V=\{v_1,v_2,v_3,v_4\}$ and the set of undirected edges $E=\{\{v_1,v_2\},\{v_2,v_3\},\{v_3,v_1\},\{v_3,v_4\}\}$ is shown in Fig. \ref{fig:graphs_examples}~(a) and an example with the same set of vertices with directed edges $E^{(d)}=\{(v_1,v_2),(v_1,v_3),(v_3,v_1),(v_3,v_2),(v_3,v_4)\}$ in Fig. \ref{fig:graphs_examples}~(b).

\begin{figure}
    \centering
    \includegraphics[width=120 pt]{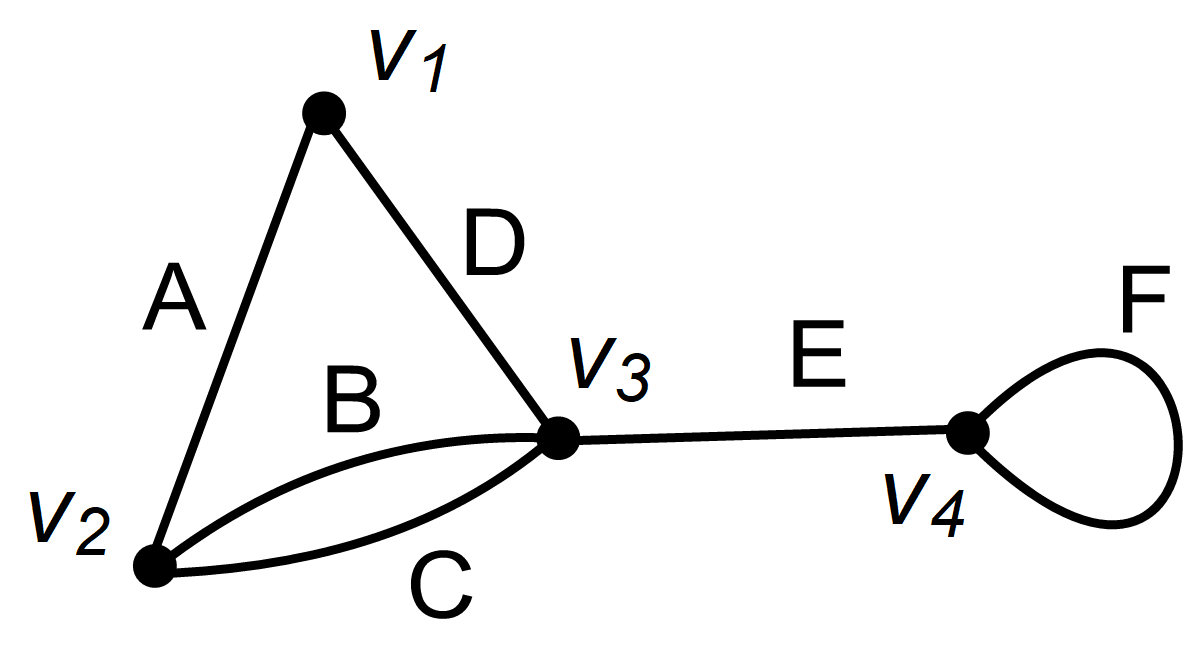}
    \caption{Example of a non-simple undirected graph.}
    \label{fig:graphs_nonsimple}
\end{figure}

Both graphs in Fig. \ref{fig:graphs_examples} are examples of so-called \textit{simple} graphs - they have no \textit{loops} and no \textit{parallel edges}. Loops are edges beginning and ending in the same vertex and parallel edges are edges connecting the same two vertices (in the same direction in the case of directed graphs). An example of a non-simple undirected graph is shown in Fig. \ref{fig:graphs_nonsimple}. We can see that for a non-simple graph it is not sufficient to denote edges by pairs of vertices and they need to be given distinct labels. In Fig. \ref{fig:graphs_nonsimple} the edge $F$ is a loop and the edges $B$ and $C$ are parallel.

An important role is assigned to graphs which can be embedded into a plane so that no edges cross.
Graphs for which this is possible are called \textit{planar}. Sometimes we call a particular embedding of a graph into a plane a \textit{plane graph}. A simple example of a non-planar graph is the complete graph (all pairs of vertices are connected by edges) on 5 vertices. Many graphs are planar including convex polyhedra like a cube or a dodecahedron. We can obtain their plane graph by extending one face and unfolding the polyhedron.
This splits the plane into simple parts without edges. They are called \textit{faces} and include the \textit{outer face} also - the rest of the plane outside the graph. Based on the number of edges enclosing a face we call the face \textit{odd-edged} or \textit{even-edged}.

An alternating sequence of vertices and edges connecting two vertices is called a \textit{walk}. If the first vertex of a walk is also the last vertex it is a \textit{closed walk}. A walk with all vertices  distinct from each other is called a \textit{path}. A closed walk where no vertex except the first/last one is repeated is called a \textit{cycle}. A walk/path/cycle is called \textit{odd} or \textit{even} based on the number of edges it covers. If a graph has no odd cycle, it is so called \textit{bipartite} - the set of vertices can be split into two subsets where there are no within-subset edges.

In an undirected graph the number of edges connected to a vertex $v\in V$ is the \textit{degree} of the vertex and is denoted $\mathrm{d}(v)$. In a directed graph a vertex has both the \textit{in-degree} and the \textit{out-degree} which are numbers of edges terminating in the vertex and originating in it respectively. In this work we only deal with directed graphs where the in-degrees and the out-degrees of all vertices are the same. Thus we only call it degree of a vertex. Note that both ends of loops contribute to (in/out) degrees of vertices of directed and undirected graphs. If the degree of all vertices is the same, the graph is \textit{regular} and in particular if the degree is $k$, the graph is called \textit{k-regular}.

In some applications we ask, how many colors are sufficient for coloring all edges of an undirected graph so that no edges of the same color meet in a common vertex. If the maximal degree of a vertex in a graph is $\Delta$, then by Vizing's theorem then the graph can be edge-colored with $\Delta+1$ colors. On the other hand, the coloring with $\Delta$ colors only exists for some graphs.

\section{Asymptotic dynamics of dynamically percolated Grover quantum walks - proofs and extensions}
\label{app:asymptotics-proofs}
In this appendix we provide proofs and details of statements used in section \ref{section_asymptotic} of the main text. Let us remind that we investigate dynamically percolated quantum walk with the Grover coin and the reflecting shift operator and shift operators with cyclic local permutations. The structure graph is assumed to be an arbitrary simple connected graph with vertices of maximal degree $3$. In the state graph unpaired loops are added so that the state graph becomes $3$-regular.

We denote three outgoing directed edges (directions) from each vertex $H$, $A$, $D$. This notation is deliberately borrowed from the honeycomb lattice - horizontal, anti-diagonal, diagonal. We denote base states by a corresponding vertex and direction symbol, e.g. $|v, H\rangle$. In each vertex, the computational basis is used in the order $H$, $A$, $D$, so explicitly
\begin{align*}
\ket{H} =
\left[
\begin{array}{c}
 1 \\
 0 \\
 0 \\
\end{array}
\right],
\ket{A} =
\left[
\begin{array}{c}
 0 \\
 1 \\
 0 \\
\end{array}
\right],
\ket{D} =
\left[
\begin{array}{c}
 0 \\
 0 \\
 1 \\
\end{array}
\right].
\end{align*}

\subsection{Independence of common eigenstate conditions}
\label{appendix_independence}

We first show that the equations (resulting from the shift and coin conditions) for the construction of common eigenstates corresponding to the eigenvalue $-1$ in the dynamically percolated reflecting Grover quantum walk on a 3-regular state graph investigated in section \ref{asymptotic_reflecting} are linearly dependent if and only if the structure graph is bipartite and there are no added loops in the state graph. After removing one freely chosen equation from the set, the remaining set becomes linearly independent as well.

In general, the common eigenstate can be written as $\ket{\psi} = \sum_{v\in V, d\in \{H,A,D\}}\psi_{v,d} \ket{v,d}$, where coefficients $\psi_{v,d}$ are determined by linear shift and coin conditions.
The equations given by the shift condition (\ref{p_shift}) are clearly linearly independent. Thus, it is sufficient to work with undirected edges (the two vector elements are the same due to the shift condition) plus the loops. Consider vertex of the first generation $v_{11}$ with its three neighboring vertices of the second generation $v_{21}$, $v_{22}$, and $v_{23}$.
Coefficients $\psi_{v,d}$ corresponding to outgoing edges from these vertices must satisfy coin coindiotions
\begin{align}
\label{first_generation}
\psi_{v_{11} H}+\psi_{v_{11} A}+\psi_{v_{11} D}=0
\end{align}
and
\begin{align*}
0 = \psi_{v_{21} H}+\psi_{v_{21} A}+\psi_{v_{21} D} = \psi_{v_{11} H}+\psi_{v_{21} A}+\psi_{v_{21} D}, \\
0 = \psi_{v_{22} H}+\psi_{v_{22} A}+\psi_{v_{22} D} = \psi_{v_{22} H}+\psi_{v_{11} A}+\psi_{v_{22} D}, \\
0 = \psi_{v_{23} H}+\psi_{v_{23} A}+\psi_{v_{23} D} = \psi_{v_{23} H}+\psi_{v_{23} A}+\psi_{v_{11} D}, \\
\end{align*}
where we use the shift condition. Employing the shift condition, we can write down analogous coin conditions for all other coefficients associated with vertices of higher generations. Now we sum up all the equations associated with the even generations of vertices and subtract all the equations associated with the odd generations (except from the first one). This is enforced by the fact that each coefficient $\psi_{v,d}$ appears exactly in two coin conditions, except coefficients corresponding to loops, which appear only in one coin condition. This sum results in the equation (\ref{first_generation}) if and only if the graph is bipartite (the bipartition of the graph corresponds to the signs of equations) and there are no loops in the state graph (every element on an unpaired loops only appears in one equation and can not cancel out).

Further, after removing one equation for a vertex or an edge, we basically have a situation analogous to a graph with unpaired loops and therefore after the removal of any equation, the remaining set of equations is linearly independent.

\section{Basis of common eigenstates for the eigenvalue -1 in the reflecting walk}
\label{appendix_basis}
In section \ref{asymptotic_reflecting} we present the construction of the common eigenstates for the eigenvalue $-1$ in the dynamically percolated Grover quantum walk with the reflecting shift operator on a planar graph with maximal degree $3$. The aim of this part is to show that the chosen set of common eigenstates truly forms linearly independent and complete basis of this eigensubspace.

Let us recall that a common eigenstate is a state vector simultaneously fulfilling the shift condition and the coin condition. The coin condition, i.e. sum of vector elements in every vertex subspace is equal to zero, is trivially fulfilled by vectors having elements of outgoing edges from one vertex equal to $1$, $-1$ and $0$ in one vertex and all the rest elements zero. We build common eigenstates as linear combinations of these one-vertex blocks by employing the shift condition, i.e. the two vector elements corresponding to paired directed edges on one undirected edge are equal. Let us start with one one-vertex block. If both the non-zero elements $+1$ and $-1$ correspond to unpaired loops, the shift condition does not impose any additional requirement and we already have a common eigenstate. On the contrary, if a non-zero element corresponds to a paired edge, the vector element associated  with the other member of the pair must have the same (non-zero) value. This enforces us to add another one-vertex block with non-zero vector elements, which extends the support of the state. In this way we continue till the process is either terminated by two loops or it is enclosed into a cycle of even length.

The arguments above lead us to build basis of common eigenstates based on walks\footnote{Here the notion walk refers to a graph object defined in appendix \ref{app:graph theory}. A walk is similar to a path but both vertices and edges can appear repeatedly in a walk.} in the structure graph and on special graph objects which we call \textit{capped walks}. A capped walk is similar to the standard graph theory walk on the undirected structure graph, but it is terminated on both ends by directed unpaired loops from the state graph - it can be thought of as an object on the mixed graph $G^{(m)}(V,E,E^{(d)}_u)$. Hence, we can associate a common eigenstate corresponding to eigenvalue $-1$  with every closed walk of even length and capped walk of arbitrary length. Its vector elements are defined by the following procedure. Starting initially with all vector elements set to zero, we choose one edge and freely add to its corresponding vector element to either $1$ or $-1$. Then we move along the walk (standard or capped) in both directions and alternately add $+1$ and $-1$ to vector elements corresponding to a particular edge from the walk till we reach terminating loops (in the case of the capped walk) or the starting edge (in the case of even edge closed walk). This approach determines not only the support of the common eigenstate but also the values of its vector elements. All four types of common eigenstates introduced in section \ref{asymptotic_reflecting} and depicted in Fig. \ref{fig:trapped_construction} can be easily obtained using this procedure. Let us show that they are sufficient to construct basis of all common eigenstates corresponding to eigenvalue $-1$.

To succeed we have to construct $N_1=2\#V-\#E$ linearly independent common eigenstates or $N_2=2\#V-\#E+1$ in case of a bipartite $3$-regular structure graph with no loops in the state graph. First we assume $3$-regular structure graphs. Employing Euler's formula for planar graphs
\begin{align*}
\#F+\#V-\#E = 2,
\end{align*}
where $\#F$ denotes the number of graph faces including the outer face, together with relation
\begin{align*}
3\#V = 2\#E,
\end{align*}
for $3$-regular graphs we get $N_1 = \#F-2$. Let further $\#F_e$ and $\#F_o$ denote the numbers of even-edged and odd-edged faces in the structure graph respectively. Consequently, we need $N_1=\#F_e+(\#F_o-1)-1$ linearly independent common eigenstates and one additional, i.e. $N_2=\#F_e-1$ linearly independent common eigenstates, if the graph is bipartite and therefore without odd-edged faces. We construct $\#F_e$ A-type common eigenstates on the even-edged faces and further $\#F_o-1$ B-type common eigenstates by connecting one chosen odd-edged face to all the others, provided there are some. If we now remove the common eigenstate which uses the outer face, we are left with just the number of common eigenstates we need.

In the next step we prove by a contradiction that these common eigenstates are linearly independent. Suppose one of the common eigenstates is a linear combination of the others. The support of the original eigenstate contains at least one face. Since every edge of a planar graph is shared by two faces, the coefficients of the linear combination for all its neighbouring faces are determined and nonzero. Similarly, also vector elements corresponding to neighboring faces of these faces must have nonzero coefficients. This continues until we reach the outer face. Since the common eigenstate whose support contains the outer face was removed, the non-zero elements on the outer edges can not be removed in the linear combination. This proves the independence of the set of common eigenstates.

Second, we consider a general planar graph with the maximal degree $3$. In order to make state graph $3$-regular, we add unpaired loops to the state graph so we have $\sum_{v\in V}d(v)+\#L=3\#V$, where $\#L$ is the number of loops in the state graph. Again using Euler's formula and relation
\begin{align*}
\sum_{v\in V}d(v) = 2\#E,
\end{align*}
we find that the number of required common eigenstates corresponding to eigenvaue $-1$ is $N=2\#V-\#E=\#F-2+\#L$.

If there are only even-edged faces in the state graph, we construct $\#F-1$ A-type common eigenstates corresponding to even-edged face cycles  (except the outer face) and add $\#L-1$ C-type common eigenstates corresponding to capped walks where one chosen loop is connected to all the other loops. The linear independence of common eigenstates corresponding to face cycles results from the same argument as before. Common eigenstates whose support contains loops are clearly linearly independent, because each of these loops (except the one) appears only in a support of one common eigenstate.

If the structure graph contains also odd-edged faces, we construct the set of common eigenstates as follows. We again use $\#F_e$ A-type common eigenstates and $\#F_o-1$ B-type common eigenstates and remove the state which uses the outer face. Additionally, we have one D-type common eigenstate for every one of the $\#L$ loops. The independence argument is the same as for the case with loops without odd-edged faces.

\section{Search for non-p-attractors}
\label{appendix_non_p}
This section is devoted to show that the attractor space of dynamically percolated Grover quantum walk with either the reflecting shift operator or a shift operator built from any combination of cyclic local permutations on an arbitrary graph with maximal degree $3$ is the linear span of p-attractors and the identity operator.

We prove this statement for each type of a shift operator separately. We stress that both proofs are done for a particular choice of local directions $H$, $A$, $D$ in all vertices. Consequently, matrix representations of the Grover coin as well as local permutations determining the shift operator are defined with respect to this particular labeling of local states at each vertex. However, one can show that local relabeling of edges at each vertex will not result in additional non-p-attractors. Indeed, if the states are reordered by some permutation $Q$, i.e. local states are transformed as $\ket{\psi}_2 = Q\ket{\psi}_1$, the corresponding modification of some operator $O_1$ is $O_2 = QO_1Q^\dagger$. Since the Grover matrix  commutes with all permutations, it is not affected by this reordering. Concerning local permutations, the reflecting shift operator has all local permutations equal to the identity, i.e. $P=I$, and hence the reordering has no effect. For shift operators built from local cyclic permutations, it can be shown, simply by taking all combinations of cyclic permutations $P$ and local permutations $Q$ on $3$-dimensional space, that the new local permutation $QPQ^\dagger$ is again a cyclic permutation. Therefore, since we consider arbitrary distributions of local rotations among vertices, the reordering just corresponds to a different case being investigated.

\subsection{Reflecting shift operator}
\label{appendix_non_p_reflecting}
The coin condition requires that any one-vertex block of a general attractor must follow the local coin equation (\ref{locally})
\begin{align}
\label{coin_equation_locally}
G_3 \Xi G_3^\dagger &= \lambda \Xi,
\end{align}
where $\Xi$ represents a general form of one-vertex blocks $X^u_v$ ($u,v\in V$) of the whole attractor $X$. Straightforward calculations reveal that for the eigenvalue $-1$, the basis of its solutions for each one-vertex block can be chosen as
\begin{align*}
\Xi^\alpha&=\left[
\begin{array}{ccc}
1 & 0 & 1 \\
0 & -1 & 0 \\
0 & -1 & 0
\end{array}
\right],
\Xi^\beta=\left[
\begin{array}{ccc}
0 & 1 & 1 \\
-1 & 0 & 0 \\
-1 & 0 & 0
\end{array}
\right], \\
\Xi^\gamma&=\left[
\begin{array}{ccc}
0 & 0 & 1 \\
-1 & -1 & 0 \\
0 & 0 & 1
\end{array}
\right],
\Xi^\delta=\left[
\begin{array}{ccc}
0 & 0 & 0 \\
1 & 1 & 1 \\
-1 & -1 & -1
\end{array}
\right].
\end{align*}
Therefore, all one-vertex blocks of a possible attractor corresponding to the eigenvalue $-1$ must follow the general form
\begin{align}
\label{eigenspace_m1}
\Xi&=\left[
\begin{array}{ccc}
\alpha & \beta & \alpha+\beta+\gamma \\
-\beta-\gamma+\delta & -\alpha-\gamma+\delta & \delta \\
-\beta-\delta & -\alpha-\delta & \gamma-\delta
\end{array}
\right].
\end{align}

Any attractor must simultaneously satisfy shift conditions (\ref{attractor_shift_condition}). Let us assume that vertices $1$ and $2$ are connected, for example, by a horizontal edge. Then the shift condition for corresponding matrix elements of the possible attractor $X$ yields
\begin{align}
\label{shift_cond_2_1}
X^{1H}_{1H}=X^{2H}_{2H},\quad X^{1H}_{2H}=X^{2H}_{1H}
\end{align}
and
\begin{align}
\label{shift_cond_2_2}
X^{1H}_{1A,1D,2A,2D}&=X^{2H}_{1A,1D,2A,2D}, \\
X^{1A,1D,2A,2D}_{1H}&=X^{1A,1D,2A,2D}_{2H}, \nonumber
\end{align}
where multiple indices are just a short-hand notation for multiple equalities. All vertex blocks $X^1_1, X^2_1, X^1_2$ and $X^2_2$ have the same form (\ref{eigenspace_m1}). Let us denote parameters of individual blocks by corresponding vertices. In terms of these parameters, the shift conditions (\ref{shift_cond_2_1}) and (\ref{shift_cond_2_2}) (with the overall signs chosen for further convenience) read
\begin{align}
\label{shift_cond_m1}
X^{1H}_{1H}=X^{2H}_{2H} &\rightarrow \alpha_{11}=\alpha_{22}, \\ \nonumber
X^{1H}_{2H}=X^{2H}_{1H} &\rightarrow \alpha_{12}=\alpha_{21}, \\ \nonumber
X^{1H}_{1A}=X^{2H}_{1A} &\rightarrow \beta_{11}=\beta_{21}, \\ \nonumber
X^{1H}_{1D}=X^{2H}_{1D} &\rightarrow \alpha_{11}+\beta_{11}+\gamma_{11}=\alpha_{21}+\beta_{21}+\gamma_{21}, \\  \nonumber
X^{1H}_{2A}=X^{2H}_{2A} &\rightarrow -\beta_{12}=-\beta_{22}, \\ \nonumber
X^{1H}_{2D}=X^{2H}_{2D} &\rightarrow -\alpha_{12}-\beta_{12}-\gamma_{12}=-\alpha_{22}-\beta_{22}-\gamma_{22}, \\ \nonumber
X^{1A}_{1H}=X^{1A}_{2H} &\rightarrow -\beta_{11}-\gamma_{11}+\delta_{11}=-\beta_{12}-\gamma_{12}+\delta_{12}, \\ \nonumber
X^{1D}_{1H}=X^{1D}_{2H} &\rightarrow -\beta_{11}-\delta_{11}=-\beta_{12}-\delta_{12}, \\ \nonumber
X^{2A}_{1H}=X^{2A}_{2H} &\rightarrow \beta_{21}+\gamma_{21}-\delta_{21}=\beta_{22}+\gamma_{22}-\delta_{22}, \\ \nonumber
X^{2D}_{1H}=X^{2D}_{2H} &\rightarrow \beta_{21}+\delta_{21}=\beta_{22}+\delta_{22}. \nonumber
\end{align}

We are looking for non-p-attractors, which additionally must violate some of the equations (\ref{broken_shift_cond}). In terms of these parameters, the condition $X^{1H}_{1H} \neq X^{1H}_{2H}$ takes the form $\alpha_{11} \neq \alpha_{12}$. However, by summing all the equations (\ref{shift_cond_m1}), we obtain the equality $\alpha_{11}=\alpha_{21}$. Thanks to the symmetry of the walk we obtain the same result for any other pair of vertices connected by an edge and so the equality (\ref{broken_shift_cond}) can not be broken. (The choice of a horizontal edge is just used for notation.) Therefore, we can conclude that there are no other attractors for the eigenvalue -1 apart from the p-attractors.\\

For the eigenvalue $1$, the basis of solutions of the local coin equation (\ref{coin_equation_locally}) can be chosen as
\begin{align*}
\Xi^\alpha&=\left[
\begin{array}{ccc}
1 & 0 & 0 \\
0 & 0 & 1 \\
0 & 1 & 0
\end{array}
\right],
\Xi^\beta=\left[
\begin{array}{ccc}
0 & 0 & 1 \\
0 & 1 & 0 \\
1 & 0 & 0
\end{array}
\right], \\
\Xi^\gamma&=\left[
\begin{array}{ccc}
0 & 1 & 0 \\
1 & 0 & 0 \\
0 & 0 & 1
\end{array}
\right],
\Xi^\delta=\left[
\begin{array}{ccc}
0 & 1 & 0 \\
0 & 0 & 1 \\
1 & 0 & 0
\end{array}
\right], \\
\Xi^\epsilon&=\left[
\begin{array}{ccc}
0 & 0 & 1 \\
1 & 0 & 0 \\
0 & 1 & 0
\end{array}
\right].
\end{align*}
Therefore all one-vertex blocks of a possible attractor corresponding to eigenvalue $1$ are of the general form
\begin{align}
\label{eigenspace}
\Xi&=\left[
\begin{array}{ccc}
\alpha & \gamma+\delta & \beta+\epsilon \\
\gamma+\epsilon & \beta & \alpha+\delta \\
\beta+\delta & \alpha+\epsilon & \gamma
\end{array}
\right].
\end{align}
Consider three vertices denoted by numbers $1,2$ and $3$. We assume, without loss of generality, that vertices 1 and 2 are connected by a horizontal edge and 2 and 3 are connected by an anti-diagonal edge. Shift conditions involving these three vertices read
\begin{align}
\label{set3}
X^{1H}_{1H}=X^{2H}_{2H}&,\quad X^{1H}_{2H}=X^{2H}_{1H}, \\
X^{2A}_{2A}=X^{3A}_{3A}&,\quad X^{2A}_{3A}=X^{3A}_{2A} \nonumber
\end{align}
and
\begin{align}
\label{set4}
X^{1H}_{1A,1D,2A,2D,3H,3A,3D}&=X^{2H}_{1A,1D,2A,2D,3H,3A,3D}, \\ \nonumber
X^{1A,1D,2A,2D,3H,3A,3D}_{1H}&=X^{1A,1D,2A,2D,3H,3A,3D}_{2H}, \\ \nonumber
X^{2A}_{2H,2D,3H,3D,1H,1A,1D}&=X^{3A}_{2H,2D,3H,3D,1H,1A,1D}, \\ \nonumber
X^{2H,2D,3H,3D,1H,1A,1D}_{2A}&=X^{2H,2D,3H,3D,1H,1A,1D}_{3A}. \nonumber
\end{align}
Employing the structure (\ref{eigenspace}) of each one-vertex block with parameters corresponding to given pairs of vertices, the equalities (\ref{set3}) can be rewritten as
\begin{align}
\label{equations1}
\alpha_{11}=\alpha_{22}&,\quad \alpha_{12}=\alpha_{21}, \\
\beta_{22}=\beta_{33}&,\quad \beta_{23}=\beta_{32}. \nonumber
\end{align}

Similarly, we rewrite a relevant part of equations (\ref{set4})
\begin{align*}
X^{1H}_{3D}=X^{2H}_{3D} &\rightarrow \beta_{13}+\epsilon_{13} = \beta_{23}+\epsilon_{23}, \\
X^{3D}_{1H}=X^{3D}_{2H} &\rightarrow \beta_{31}+\delta_{31} = \beta_{32}+\delta_{32}, \\
X^{1H}_{2D}=X^{2H}_{2D} &\rightarrow -\beta_{12}-\epsilon_{12} = -\beta_{22}-\epsilon_{22}, \\
X^{2D}_{1H}=X^{2D}_{2H} &\rightarrow -\beta_{21}-\delta_{21} = -\beta_{22}-\delta_{22}, \\
X^{2D}_{2A}=X^{2D}_{3A} &\rightarrow -\alpha_{22}-\epsilon_{22} = -\alpha_{23}-\epsilon_{23}, \\
X^{1D}_{2A}=X^{1D}_{3A} &\rightarrow \alpha_{12}+\epsilon_{12} = \alpha_{13}+\epsilon_{13}, \\
X^{2A}_{1D}=X^{3A}_{1D} &\rightarrow \alpha_{21}+\delta_{21} = \alpha_{31}+\delta_{31}, \\
X^{2A}_{2D}=X^{3A}_{2D} &\rightarrow -\alpha_{22}-\delta_{22} = -\alpha_{32}-\delta_{32}, \\
X^{2A}_{1A}=X^{3A}_{1A} &\rightarrow \beta_{21} = \beta_{31}, \\
X^{1A}_{2A}=X^{1A}_{3A} &\rightarrow \beta_{12} = \beta_{13}, \\
X^{1H}_{3H}=X^{2H}_{3H} &\rightarrow \alpha_{13} = \alpha_{23}, \\
X^{3H}_{1H}=X^{3H}_{2H} &\rightarrow \alpha_{31} = \alpha_{32}.
\end{align*}
By summing all these equations and using (\ref{equations1}) we finally obtain
\begin{align}
\label{no_non_p}
\alpha_{12}-\alpha_{22}&=\beta_{23}-\beta_{22}.
\end{align}
The actual choice of horizontal and anti-diagonal connecting edges is irrelevant. A different choice of connecting edges will simply replace parameters $\alpha$ and $\beta$ in equation (\ref{no_non_p}) by parameters corresponding to the new choice. As the structure graph is connected, we conclude, that if the equality (\ref{broken_shift_cond}) holds for one pair of connected vertices (for one edge), it holds for the whole attractor, and so it is a p-attractor. 

Based on this finding we can show that the attractor space associated with eigenvalue 1 is the linear span of p-attractors and the identity operator $I$. Suppose there is another attractor $X$ corresponding to the eigenvalue 1 and let us choose two vertices $v_1$ and $v_2$ (for the sake of notation let $v_1$ and $v_2$ be connected, for example, by a horizontal edge). There clearly exist complex numbers $z_1$ and $z_2$ such that for the attractor $Y = z_1 I + z_2 X$ the equation $\alpha^{Y}_{12}=z_1 \alpha^{I}_{12} + z_2 \alpha^{X}_{12} = z_1 \alpha^{I}_{22} + z_2 \alpha^{X}_{22}=\alpha^{Y}_{22}$ holds. From the above paragraph, it follows that the attractor $Y$ is a p-attractor. Hence, an arbitrary attractor is a linear combination of the identity attractor and a p-attractor. We conclude, that there are no other linearly independent non-p-attractors apart from the identity operator.

\subsection{Shift operators with cyclic local permutations}
\label{appendix_non_p_cyclic}
In this part we investigate an existence of non-p-attractors for percolated quantum walks on graphs with maximal degree $3$ equipped with a shift operator built from cyclic local permutations. Assuming a general distribution of these local permutations, the local coin equation (\ref{locally}) can take four different forms
\begin{align*}
G_3 P^{CW}_{v_1} \Xi (G_3 P^{CW}_{v_2})^\dagger &= \lambda \Xi, \\
G_3 P^{CW}_{v_1} \Xi (G_3 P^{CCW}_{v_2})^\dagger &= \lambda \Xi, \\
G_3 P^{CCW}_{v_1} \Xi (G_3 P^{CW}_{v_2})^\dagger &= \lambda \Xi, \\
G_3 P^{CCW}_{v_1} \Xi (G_3 P^{CCW}_{v_2})^\dagger &= \lambda \Xi,
\end{align*}
where $\Xi$ represents general form of blocks $X^u_v$ ($u,v\in V$) of the whole attractor $X$.

In all four cases, the set of eigenvalues $\{1,\e^{i\frac{2}{3}\pi},\e^{-i\frac{2}{3}\pi},\e^{i\frac{\pi}{3}},\e^{-i\frac{\pi}{3}}\}$  and dimensions of their eigenspaces are the same, but their corresponding eigenvector spaces differ in general.
To simplify notation we denote $\omega=\e^{i\frac{\pi}{3}}$. We start our analysis of non-p-attractors with the eigenvalue $\lambda=\omega^2$ and the CW cyclic permutation in both vertices $v_1, v_2$ connected by a horizontal edge. The corresponding eigenspace is one-dimensional and a straightforward calculation reveals that the one-vertex block has the general form
\begin{align*}
\Xi^{CW}_{CW} &= \left[
\begin{array}{ccc}
 \alpha & \alpha\omega^{-2} & \alpha\omega^{2} \\
 \alpha\omega^{-2} & \alpha\omega^{2} & \alpha \\
 \alpha\omega^{2} & \alpha & \alpha\omega^{-2}
\end{array}
\right].
\end{align*}
The shift condition $X^{1A}_{1H}=X^{1A}_{2H}$ reads $\omega^{-2}\alpha_{11}=\omega^{-2}\alpha_{12} \Rightarrow \alpha_{11}=\alpha_{12}$, which in turn implies $X^{1H}_{1H}=X^{1H}_{2H}$. Thus equality (\ref{broken_shift_cond}) holds and there are no non-p-attractors.
If we have CW local permutation in vertex $v_1$ and CCW permutation in vertex $v_2$, the general form of block $X^1_2$ is
\begin{align*}
\Xi^{CW}_{CCW} &= \left[
\begin{array}{ccc}
 \alpha & \alpha\omega^{2} & \alpha\omega^{-2} \\
 \alpha\omega^{-2} & \alpha & \alpha\omega^{2} \\
 \alpha\omega^{2} & \alpha\omega^{-2} & \alpha
\end{array}
\right].
\end{align*}
The same reasoning shows again that there is no non-p-attractor. Analogously, the same conclusion holds for the situation with the CCW local permutation in both vertices and for all three cases for the eigenvalue $\lambda=\omega^{-2}$.


For the eigenvalue $\lambda=\omega$ the eigensubspace is always two-dimensional. If there is the CW cyclic permutation in both vertices $v_1, v_2$, the one vertex block is of the form
\begin{align*}
\Xi^{CW}_{CW} &= \left[
\begin{array}{ccc}
 \alpha & \alpha\omega^{-1}-\beta & \beta\omega^{2} \\
 \beta & \alpha\omega^{-2} & -\alpha+\beta\omega \\
 \alpha\omega+\beta\omega^{-1} & \beta\omega^{-2} & \alpha\omega^{2}
\end{array}
\right].
\end{align*}
Without loss of generality we again assume that $v_1$ and $v_2$ are connected by a horizontal edge. The shift condition $X^{1A}_{1H}=X^{1A}_{2H}$ implies  $\beta_{11}=\beta_{12}$ and $X^{1D}_{1H}=X^{1D}_{2H}$ implies $\omega\alpha_{11}+\omega^{-1}\beta_{11}=\omega\alpha_{12}+\omega^{-1}\beta_{12}$ and hence $\alpha_{11}=\alpha_{12}$. This means that $X^{1H}_{1H}=X^{1H}_{2H}$, the equality (\ref{broken_shift_cond}) holds and we prove that in this case there are no non-p-attractors. If we have the CW permutation in $v_1$ and the CCW permutation in $v_2$, the general form of the one-vertex block reads
\begin{align*}
\Xi^{CW}_{CCW} &= \left[
\begin{array}{ccc}
 \alpha & \beta\omega^{2} & \alpha\omega^{-1}-\beta \\
 \beta & -\alpha+\beta\omega & \alpha\omega^{-2} \\
 \alpha\omega+\beta\omega^{-1} & \alpha\omega^{2} & \beta\omega^{-2}
\end{array}
\right].
\end{align*}
and we again find that there is no non-p-attractor. The same statement holds for the CCW permutation in both vertices and for all three cases if $\lambda=\omega^{-1}$.

For the eigenvalue $\lambda=1$, 
the general form of one-vertex block
\begin{align*}
\Xi = \left[
\begin{array}{ccc}
 \alpha & \gamma & \beta \\
 \beta & \alpha & \gamma \\
 \gamma & \beta & \alpha
\end{array}
\right]
\end{align*}
is the same for all combinations of CW and CCW local permutations. Similarly as for the reflecting shift operator, we assume three vertices,  where vertices 1 and 2 are connected by a horizontal edge and vertices 2 and 3 by an anti-diagonal edge. The shift condition (\ref{set3}) gives $\alpha_{11}=\alpha_{22}=\alpha_{33}$, $\alpha_{12}=\alpha_{21}$ and $\alpha_{23}=\alpha_{32}$. Equation (\ref{set4}) $X^{2A}_{1A}=X^{3A}_{1A}$ implies $\alpha_{21}=\alpha_{31}$ and $X^{3H}_{1H}=X^{3H}_{2H}$ gives $\alpha_{31}=\alpha_{32}$. A combination of these equalities results in $\alpha_{11}-\alpha_{12} = \alpha_{22}-\alpha_{23}$. Using the same arguments as for the reflecting shift operator, one can show that all attractors can be obtained as linear combinations of p-attractors and to the identity operator.

\end{document}